\documentclass[authoryear]{elsarticle}
\usepackage[a4paper, total={6in, 10in}]{geometry}
\usepackage{booktabs}
\usepackage{float}
\usepackage{bbold}
\usepackage{amsmath}
\usepackage{amsthm}
\usepackage{amsmath,amssymb,amsfonts,mathrsfs,hyperref,microtype, amsthm}
\usepackage{graphicx}
\usepackage{graphicx,mathabx}
\usepackage{enumitem}  
\graphicspath{ {images/} }
\numberwithin{equation}{section}

\newtheorem{definition}{Definition}[section]
\newtheorem{example}{Example}
\newtheorem*{example*}{Example}

\newcommand{\BS}{\boldsymbol}

\newcommand{\Rmnum}[1]{\expandafter\@slowromancap\romannumeral #1@}

\journal{}
\makeatletter
\def\ps@pprintTitle{%
   \let\@oddhead\@empty
   \let\@evenhead\@empty
   \def\@oddfoot{\reset@font\hfil\thepage\hfil}
   \let\@evenfoot\@oddfoot
}

\begin{document}

\begin{frontmatter}
\author[a]{Helmi Shat\corref{cor1}}
\ead{hshat@ovgu.de}
\author[a]{Norbert Gaffke}
\ead{norbert.gaffke@ovgu.de}
\address[a]{\small Institute for Mathematical Stochastics, Otto-von-Guericke University Magdeburg, \\ \small PF 4120,
39016 Magdeburg, Germany }

 \title{ Optimal Accelerated
Degradation Testing Based on Bivariate Gamma Process with Dependent Components}

\begin{abstract}
Accelerated degradation testing (ADT) is one of the major approaches in reliability engineering which allows
accurate estimation of reliability characteristics of highly reliable systems within a relatively short time. 
The testing data are extrapolated through a physically reasonable statistical model to obtain estimates of lifetime quantiles 
at normal use conditions.
The Gamma process is a natural model for degradation, which exhibits a monotone and strictly increasing degradation path.
In this work, optimal experimental designs are derived for ADT with two response components. 
We consider the situations of independent as well as dependent marginal responses where the observational times 
are assumed to be fixed and known. The marginal degradation paths are assumed to follow a Gamma process 
where a copula function is utilized to express the dependence between both components. 
For the case of independent response components the optimal design minimizes the asymptotic variance 
of an estimated  quantile of the failure time distribution at the normal use conditions.
For the case of dependent response components the $D$-criterion is adopted to derive $D$-optimal designs. 
Further, $D$- and $c$-optimal designs are developed when the copula-based models are reduced to  
bivariate binary outcomes.  
\end{abstract}


\end{frontmatter}

\section{Introduction}
In recent years, the increasing demand for highly reliable products has motivated a noticeable growth of research interest 
in the area of degradation testing. For systems with high reliability, it is time consuming to do reliability assessment 
based on traditional degradation tests. Hence, ADT ensures an efficient reliability and life time assessment within 
relatively short testing times by statistically extrapolating the obtained actual degradation data. 
In fact, the majority of research on ADT has considered the case of one performance characteristic or the case of multiple 
but independent failure modes. For example, \citep{1603894} used a stochastic diffusion process to model a typical
 step stress ADT problem with a single failure mode under the constraint that the total experimental cost does not 
exceed a predetermined budget. The optimal settings of the design variables were obtained by minimizing the asymptotic 
variance of an estimated quantile of the product's lifetime distribution.
\citep{huang2003reliability} presented a reliability analysis of electronic devices
with independent competing failure modes involving performance aging degradation. The authors used  Weibull distributions 
to describe the time-to-failure of a catastrophic failure mode and that of a degradation failure mode. 
\citep{bai1991optimum} introduced optimal simple step stress for products with competing causes of failure, 
where the life times for the different failure causes were assumed to be independent and exponentially distributed.
The authors presented optimal plans which minimize the sum over all failure causes of asymptotic variances of 
the estimated log mean lives at design stress. Modern products usually have complex structure with multiple failure 
mechanisms as well as multiple degradation measures. Thus, it is realistic to assume some kind of dependence among 
different failure components. In the past decade, copula-based modelling has become an efficient tool in many areas 
of applied statistics, see \citep{aghakouchak2010copula} and \citep{embrechts2001modelling}. 
For instance, \citep{perrone2016optimal} has provided an equivalence theorem for binary bivariate copula models  
that allows applications of efficient design algorithms and quick checks of whether a design is optimal 
or at least efficient. With an application in cancer
clinical trials, \citep{yin2009bayesian} proposed a Bayesian adaptive design
for dose finding that is based on a Clayton copula model to account for the synergistic effect of two or more drugs 
in combination. Considering modern complex system, Levy stochastic processes, i.e. Gamma process,
Wiener process (\citep{lim2011optimal} and \citep{xiao2016optimal}), and Inverse Gaussian
process (\citep{peng2014bayesian} and \citep{ye2014accelerated}), were used to model the degradation path. 
For instance, \citep{tsai2012optimal} and \citep{amini2016optimal} discuss the problem of optimal design for 
degradation testing based on a Gamma degradation process with random effects. \citep{tsai2012optimal} 
considered several decision variables such as the sample
size, inspection frequency, and measurement numbers in order to find the $c$-optimal decision variables. 
\citep{doi:10.1080/03610926.2018.1459718}  addressed the optimal design problems for constant stress
ADT based on Gamma processes with fixed effects and random effects. For $D$-, $V$- and $A$-optimality criteria, the authors proved
that optimal constant stress ADT plans with multiple stress only use the minimum and the maximum stress levels. 

The Archimedean, Clayton, Frank and Gumbel copulas are intensively used to describe the dependence among 
different failure components when the marginal degradation paths correspond to Levy stochastic processes, 
see \citep{mireh2019copula}.
For example, \citep{zhou2010bivariate} and  \citep{guo2017bivariate} followed a similar approach 
through considering a system with multiple failure components where the marginal degradation paths are governed 
by Gamma processes. They utilized the Frank copula to describe the dependence of failure components. 
Furthermore, the authors used the Bayesian MCMC method in order to efficiently evaluate the maximum likelihood estimator. 
In addition, \citep{adegbola2019multivariate} proposed a multivariate Gamma process to model dependent deterioration 
phenomena that collectively define the service life of infrastructure assets. 
\citep {liu2014reliability} developed a reliability model for systems with s-dependent degradation processes 
using several Archimedean copulas. The marginal degradation processes were assumed to be 
inverse Gaussian with a time scale transformation. Furthermore, the authors incorporated a random drift 
to account for a possible heterogeneity in population, with an application to fit the crack length growth problem. 
Considering a Wiener process, \citep{pan2013bivariate} and \citep{doi:10.1080/03610918.2010.534227} presented a 
bivariate stochastic process where the dependence of the performance characteristics
were described by a Frank copula. In addition, the authors used MCMC to jointly estimate the parameters of the two 
performance characteristics as well as the parameter of the Frank copula. In order to provide a more flexible 
dependence structure between competing failure modes, \citep{wang2011modeling} introduced time-varying copulas 
to develop an $s$-dependent competing risk model for systems subject to multiple degradation processes and random shocks.
Moreover, \citep{doi:10.1177/1748006X13481928} investigated the effect
of various copulas for modeling dependence structures between variables on reliability under incomplete information.
The authors formulated a reliability problem and a direct integration method for calculating the probability of failure. 
\citep{mercier2012bivariate} discussed the intervention scheduling of a railway track, based on the observation of two dependent
randomly increasing degradation components. The authors used trivariate reduction for constructing a bivariate 
Gamma process that describes the dependency between the two components. Further, they utilized an EM-algorithm 
to compute the maximum likelihood estimators of the model parameters. 
In regards to ALT, 
\citep{hove2017estimation} 
utilized the Frank copula to model the general dependence 
structure between the conceptual lifetimes of system with multiple competing risks. With an application to finance, 
\citep{semeraro2008multivariate} proposed a generalized bivariate variance Gamma process by subordinating a 
multivariate Wiener process with independent components by a multivariate Gamma subordinator. 
With an application to toxicity trials, \citep{denman2011design} derived locally $D$-optimal designs 
for dependent bivariate binary data, where several Archimedean, i.e. Clayton, Frank and Gumbel, copulas were utilized 
to describe the dependence among the marginal regression models. 
Further, \citep{mireh2019copula} proposed a simulation-based reliability analysis for systems with dependent Gamma
degradation processes and Weibull distributed hard failure times. The authors used the Frank copula to represent 
the dependence between failure modes.
\citep{pan2016model} introduced a copula based bias correction approach to address model uncertainty in a defined 
product design. In addition, the results were illustrated by a modified vehicle side impact response
case study.\\

The rest of the present paper is organized as follows. In Section \ref{indepebendebtcgpulasectionw5345jhg} we obtain an optimal 
experimental design for a bivariate Gamma model with independent marginal components. In Section \ref{dependentcpopulasection345}  
we develop $D$-optimal designs for bivariate Gamma models with dependent responses based on the 
Frank copula function or the Gaussian copula function, respectively. Section \ref{bivariatewithcorrelation} introduces $D$- and $c$-optimal designs 
for ADT with dependent failure modes when the copula-based model is reduced to bivariate binary outcomes.  The numerical computations were made by using the R programming language\citep{koss}.

\section{Bivariate Gamma process with independent components}
\label{indepebendebtcgpulasectionw5345jhg}
\subsection {Model construction}
\label{indepebendebtcopulasectionw5345jhgfdsfsfdsf}

The Gamma process is a natural stochastic model for degradation processes in which degradation occurs gradually over time in a sequence of 
independent increments. 
In this section, we assume that the testing unit has two failure modes where the marginal degradation paths are given by Gamma processes  
in terms of a standardized continuous time variable $t\ge0$, and the two marginal Gamma processes are independent.
It is further assumed that each of the marginal (standardized) stress levels $x_l,\,l=1,2$, is a scalar in the 
standardized interval $[0,1]$. 
The joint stress variable $\mathbf{x}=(x_1,x_2)$ can be chosen by the experimenter from the experimental region 
$\mathcal{{X}}=[0,1]^2$. 
Below we clarify the approximation of the Gamma model with a generalized linear model approach.
For (locally) optimal design, the information matrix as a function of $\mathbf{x}$ (at given values of the model parameters) 
is of basic interest and will be derived in Subsection \ref{informationnnnnnn1243sdf}.  Locally $c$-optimal designs will be presented in Subsection \ref{optimalityyyyyzzzrt56},
where the particular $c$-criterion expresses the asymptotic variance of an estimated quantile of the failure time distribution.  

A Gamma process $Z_t^{(l)}$, $t\ge0$,  considering the response component $l=1,2$ 
is a stochastic process with independent and Gamma distributed increments. A degradation increment $Z_t^{(l)}-Z_s^{(l)}$, 
$0\le s< t$, is Gamma distributed with shape parameter $\gamma_l\cdot(t-s)$ and scale parameter $\nu_l$. The scale parameter $\nu_l$
is a known positive constant, while the shape rate  $\gamma_l$ is a positive function of the stress variable $x_l$ 
and some further model parameters (see below). The two marginal processes  $Z_t^{(1)}$, $t\ge0$, and $Z_t^{(2)}$, $t\ge0$,
are assumed to be independent.
The bivariate degradation process $\BS{Z}_t=\bigl(Z_t^{(1)},Z_t^{(2)}\bigr)$ is observed at  
$k$ subsequent time points $t_j$, $j=1,\ldots,,k$, $0<t_1<\ldots<t_k$, which are prescribed in advance. 
Equivalently, the bivariate increments $\mathbf{Y}_j=\bigl(Y_j^{(1)},Y_j^{(2)}\bigr)$, $j=1,\ldots,k$, are observed,
where $Y_j^{(l)}=Z_{t_j}^{(l)}-Z_{t_{j-1}}^{(l)}$, $l=1,2$, and $t_0=0$. By the above assumptions, the bivariate increments are independent,
and the components $Y_j^{(1)}$ and $Y_j^{(2)}$ are independent for each $j$. The density of a marginal increment $Y_j^{(\ell)}$ is given by    
\begin{equation}
	\label{marginal}
	f_{jl}(y_{jl}) = \frac{y_{jl}^{\gamma_l \Delta_j - 1} e^{-y_{jl}/\nu_l}}{\gamma\big(\gamma_l \Delta_j\big) \nu_l^{\gamma_l \Delta_j}},\quad 
y_{jl}\in(\,0\,,\,\infty),
\end{equation}
where $\Delta_j=t_j-t_{j-1}$ and $\Gamma(u) = \int_0^\infty x^{u - 1} e^{ - x} \mathrm{d}x$, $u>0$, 
is the complete Gamma function, see \citep{qi2004complete}. 
In accordance with the work of \citep{shat2019experimental} for the univariate Gamma process, for the marginal shape rate $\gamma_l={\gamma}_l(x_l)$ as a function of the stress variable we consider the particular case 
\begin{equation} 
\label{shape-marginal_l}
\gamma_l(x_l) = \exp(\beta_{1l} + \beta_{2l} x_l)
\end{equation}
where the intercept and slope parameters $\beta_{1l}$ and $\beta_{2l}$ are to be estimated. Hence,   
the mean of a marginal increment is given by 
\begin{equation} 
	\label{expected-value-gamma-marginal l}
	\mu_{jl}(x_{l}) = \mathrm{E}(Y_{jl}) = \gamma_l(x_{l}) \Delta_j \nu_l = \exp(\beta_{1l} + \beta_{2l} x_{l}) \Delta_j \nu_l.
\end{equation}
Thus the mean is linked to the linear predictor $\beta_{1l} + \beta_{2l} x_l$ by the (non-linear) log link, 
and the present model is related to a generalized linear model with Gamma distributed response variables.

When an accelerated degradation test is run under a stress setting $\mathbf{x}=(x_1,x_2)$, measurements of the bivariate degradation process
at the prescribed time points $t_j$, $j=1,\ldots,k$, are made. So the increments $\mathbf{Y}_j=(y_{j1},y_{j2})$, $j=1,\ldots,k$, $l=1,2$, of the
bivariate degradation path are obtained, which follow the model of independent bivariate random variables 
$\mathbf{Y}_j=(Y_{j1},Y_{jl})$, $j=1,\ldots,k$,   
with Gamma distributed components $Y_{jl}$  according to (\ref{marginal}) and (\ref{shape-marginal_l}).
Thus, under the stress setting $\mathbf{x}$ and given the incremental data $\BS{y}=(\BS{y}_1,\ldots,\BS{y}_k)$, the 
log-likelihood of the parameter vector $\BS{\beta}=\bigl(\BS{\beta}_1^{T},\BS{\beta}_2^{T}\bigr)^{T}$, 
where $\BS{\beta}_l=(\beta_{1l},\beta_{2l})^{T}$, $l=1,2$, is given by
\begin{eqnarray}
&&\ell(\BS{\beta};\,\mathbf{x},\BS{y}) = \sum_{j=1}^k\sum_{l=1}^2 \exp\bigl(f_{jl}(y_{jl})\bigr)\label{eq-loglik-single-marginal -l}\\
&&\phantom{xxxxx}=\sum_{j=1}^k\sum_{l=1}^2\Bigl[\big(\gamma_l(x_l)\, \Delta_j - 1\big) \exp(y_{jl})  
- \frac{y_{jl}}{\nu_l} - \exp\big(\Gamma(\gamma_l(x_l)\,\Delta_j)\big) - \gamma_l(x_l)\, \Delta_j \exp(\nu_l),\nonumber\\
&&\mbox{ where  }\ \gamma_l(x_l)=\exp(\beta_{1l}+\beta_{2l}x_l),\ \ l=1,2.\nonumber  
\end{eqnarray}    
Usually, an accelerated degradation test is conducted at $n$ distinct testing units $i=1,\ldots,n$ at stress settings 
$\mathbf{x}_1,\ldots,\mathbf{x_n}$, respectively.
Note that the stress settings $\mathbf{x}_i$, $i=1,\ldots,n$ may not all be distinct. Under the assumption of independence of the testing units,
the joint log-likelihood equals the sum of the log-likelihoods over the units,
\[
\ell(\BS{\beta};\,\mathbf{x}_1,\ldots,\mathbf{x}_n,\BS{y}_1,\ldots,\BS{y_n}) = \sum_{i=1}^n\ell(\BS{\beta};\,\mathbf{x}_i,\BS{y}_i).
\]
The collection $\mathbf{x}_1,\ldots,\mathbf{x}_n$ constitutes the experimental design of the test. Since the ordering of the design points 
$\mathbf{x}_i$ (along with the response vector $\BS{y}_i$) is of no importance, a design is usually described by the set of 
of {\em distinct} points $\mathbf{x}_1',\ldots,\mathbf{x}_m'$ among the collection $\mathbf{x}_1,\ldots,\mathbf{x}_n$
and the corresponding frequencies $n_1,\ldots,n_m$ of their occurrence among $\mathbf{x}_1,\ldots,\mathbf{x}_n$.
In optimal design theory, when the sample size $n$ is kept fixed, it has become standard to use the {\em relative} 
frequencies $\omega_j=n_j/n$, $j=1,\ldots,m$, and defining an exact design for sample size $n$ by
\begin{equation}
\xi_n\,=\,\left(\begin{array}{ccc}\mathbf{x}_1' & \cdots & \mathbf{x}_m'\\ \omega_1 & \cdots & \omega_m\end{array}\right),
\label{exact-design}
\end{equation}
where $m\in\mathbb{N}$, $\mathbf{x}_1',\ldots,\mathbf{x}_m'\in{\cal X}$, and $\omega_1,\ldots,\omega_m$ are positive integer 
multiples of $1/n$ with $\sum_{j=1}^m\omega_j=1$. 
Note that the positive integer $m$, called the support size of $\xi_n$, may vary with the design.
As a mathematical relaxation one dispenses the discrete character of the weights, allowing 
any positive weights $\omega_j>0$, $j=1,\ldots,m$, with $\sum_{j=1}^m\omega_j=1$. Then the r.h.s. of
(\ref{exact-design}) defines an {\em approximate} design, for short:  a design $\xi$. 
The weight $w_j$ given by $\xi$ to the support point $\mathbf{x}_j'$ will also be denoted by $\xi(\mathbf{x}_j)$. In what follows, we 
will employ the approximate design theory for deriving optimal designs, (see e. g. \citep{silvey1980optimal})..

\subsection{Information matrix}
\label{informationnnnnnn1243sdf}
By the log-likelihood $\ell(\BS{\beta};\mathbf{x},\BS{y})$ from (\ref{eq-loglik-single-marginal -l}) 
the elemental Fisher information matrix of $\mathbf{x}$ at $\BS{\beta}$ is given by either of following two representations,
\begin{equation}
\mathbf{M}(\mathbf{x},\BS{\beta}) = {\rm E}\left(\Bigl[\frac{\partial\ell(\BS{\beta}; \mathbf{x},\BS{Y})}{
\partial\BS{\beta}}\Bigr]\,\Bigl[\frac{\partial\ell(\BS{\beta}; \mathbf{x},\BS{Y})}{
\partial\BS{\beta}}\Bigr]^{T}\right) = 
-{\rm E}\Bigl(\frac{\partial^2\ell(\BS{\beta}; \mathbf{x},\BS{Y})}{
\partial\BS{\beta}\partial\BS{\beta}^{T}}\Bigr). \label{elem-info}
\end{equation}
Using the latter representation, direct calculations and observing(\ref{shape-marginal_l}) yield a block structure because of independence of the components as
\begin{eqnarray}
&&\mathbf{M}(\mathbf{x},\BS{\beta})=\left( 
\begin{array}{c@{}c@{}c}
 \mathbf{M}_1(x_1,\BS{\beta}_1) && \mathbf{0}\\
\mathbf{0} &  &  \mathbf{M}_2(x_2,\BS{\beta}_2) \\
\end{array}\right),\label{elem-info-ex_a}\\
&&\mbox{ where } \ \mathbf{x}=(x_1,x_2),\ \ \BS{\beta}_l=(\beta_{1l},\beta_{2l})^{T},\ l=1,2,\ \mbox{ and}\nonumber\\ 
&&\mathbf{M}_l(x_l,\BS{\beta}_l)=\lambda_l(x_l,\BS{\beta}_l)\left(
	\begin{array}{cc}
		1 & x_{l}
		\\
		x_{l} & x_{l}^2
	\end{array}
	\right), \ \ l=1,2, \label{elem-info-ex_b}\\
&&
\mbox{ with }\lambda_l(x_l,\BS{\beta}_l)=\gamma_l^2(x_l)\sum_{j=1}^k \Delta_j^2\psi_1\bigl(\gamma_l(x_l)\,\Delta_j\bigr),\nonumber
\end{eqnarray}
where $\psi_1$ denotes the Tri-Gamma function, i.e., 
$\psi_1(z)={\rm d}^2 \exp\Gamma(z)\big/{\rm d}z^2$, $z>0$.
As usual in the approximate design theory, for any (approximate) design 
\[
\xi=\left(\begin{array}{ccc}\mathbf{x}_1 & \cdots & \mathbf{x}_m\\ \omega_1 & \cdots & \omega_m\end{array}\right),
\]
where $m\in\mathbb{N}$, $\mathbf{x}_i\in{\cal X}$, $\omega_i>0$, $1\le i\le m$, and $\sum_{i=1}^m\omega_i=1$,
the information matrix of $\xi$ at a parameter point $\BS{\beta}$ is given by
\[
\mathbf{M}(\xi,\BS{\beta})=\sum_{i=1}^m\omega_i\mathbf{M}(\mathbf{x}_i,\BS{\beta}).
\]
By the block-diagonal structure of the elemental information matrices (\ref{elem-info-ex_a}),
the information matrix of $\xi$ is again block-diagonal where the blocks are given by the information matrices
of the marginal designs w.r.t. the marginal models,
\begin{eqnarray}
&&\textbf{M}({\xi},\BS{\beta})=\left( 
\begin{array}{c@{}c@{}c}
 \mathbf{M}_1(\xi_1,\BS{\beta}_1) & & \mathbf{0}\\
\mathbf{0}   & & \mathbf{M}_2(\xi_2,\BS{\beta}_2) \end{array}\right).\label{infomatrix-xi}\\
&&\mbox{ where }\ \mathbf{M}_l(\xi_l,\BS{\beta}_l)=\sum_{i=1}^m\omega_i\mathbf{M}_l(x_{il},\BS{\beta}_l),\ \ l=1,2.
\label{infomatrix-xi_l}
\end{eqnarray}
Recall that ${\cal X}=[\,0\,,\,1\,]^2$, hence $\mathbf{x}_i=(x_{i1},x_{i2})$ with $x_{1i},x_{2i}\in[\,0\,,\,1\,]$,
$i=1,\ldots,m$. The designs $\xi_1$ and $\xi_2$ on $[\,0\,,\,1\,]$ are the marginal designs of $\xi$, which are defined as the projections on the corresponding components (in a measure theoretic sense).

\subsection{Optimality criterion based on failure time distribution}
\label{optimalityyyyyzzzrt56}
In ADT one considers some characteristics of the failure time distribution due to degradation under 
normal use condition $\mathbf x_u=({x}_{u1},{x}_{u2})$. Note that typically the normal use conditions $x_{u1}$ and $x_{u2}$ are outside the
normalized interval $[0,1]$ of the possible stress values $x_1$ and $x_2$ in ADT. Usually, one has $x_{ul}<0$, $l=1,2$. It is assumed that the marginal Gamma process $Z_{u,t}^{(l)}$ describing the degradation under normal use condition ${x}_{ul}$ 
has the rate $\gamma_l({x}_{ul}) = \exp(\beta_{1l} + \beta_{2l} {x}_{ul})$ according to (\ref{shape-marginal_l}) and scale $\nu_l$.  
A soft failure due to degradation is defined by exceedance of the marginal degradation paths over some failure thresholds. 
The marginal failure time $T_l$ under normal use condition $x_{ul}$ is expressed as the first time $t$ the degradation path $Z_{u,t}^{(l)}$ 
reaches or exceeds a given positive value $z_{l0}$, i.\,e.
\begin{equation}
T_l = \inf \{t \geq 0;\, Z_{u,t}^{(l)} \geq z_{l0}\}. \label{marg-failure-time}
\end{equation} 
Its distribution function is given by $F_{T_l}(t) = \mathrm{P}(T_l \leq t)$, $t\ge0$. 
By (\ref{marg-failure-time}), $T_l \leq t$ if and only if $Z_{u,t}^{(l)} \geq z_{l0}$, hence 
\begin{equation}
\label{eq-failure-time-distribution-margiunal-l}
\begin{split}
F_{T_l}(t) & = \mathrm{P}(Z_{u,t}^{(l)} \geq z_{l0}) \\
	& = \frac{1}{\Gamma(\gamma_l(x_{ul}) t)} \int_{z_{l0}}^{\infty} (z_l / \nu_l)^{\gamma_l(x_{ul}) t - 1} e^{ - z_l / \nu} \nu_l^{ - 1} \mathrm{d}z_l \\
	& = Q(\gamma_l(x_{ul}) t , z_{l0} / \nu_l)
\end{split}
\end{equation}
where $Q(s, z) = \Gamma(s, z)/\Gamma(s)$ is the regularized Gamma function and $\Gamma(s , z) = \int_{z}^{\infty} x^{s - 1} e^{ - x} \mathrm{d}x$ 
is the upper incomplete Gamma function, see \citep{WANG2015172}.

As opposed to \citep{shat2019experimental} we assume a parallel system, that is, the system fails as soon as both marginal components have failed. Denote by $T$ the joint failure time,  
$T=\max\{T_1,T_2\}$. By independence of the components its distribution function is given by
\begin{equation} 
\label{distributionnnnnnnfunction}
\begin{split} 
F_{T}(t)&= \mathrm{P}(Z_{u,t}^{(1)} \geq z_{10},Z_{u,t}^{(2)} \geq z_{20})
 \\
& =Q(\gamma_1(x_{u1}) t , z_{10} / \nu_1)\,Q(\gamma_2(x_{u2}) t , z_{20} / \nu_2).\\
\end{split}
\end{equation}
For a given $\alpha\in(\,0\,,\,1\,)$ let $t_\alpha$ be the $\alpha$-quantile of the failure time distribution of $T$, that is 
$F_T(t_\alpha) = \alpha$.
This quantile represents the time up to which on the average, under normal use conditions, 
$\alpha \cdot 100$ percent of the testing units will fail 
and $(1 - \alpha) \cdot 100$ percent of the units will persist. 
It is worth noting that the distribution function $F_{T}$ as well as its quantile
$t_\alpha$ depend on the parameter vector $\BS{\beta}$, though not expressed by our notations. 
The performance of the maximum likelood estimator $\widehat{t}_\alpha$ is measured by its asymptotic variance 
$\mathrm{aVar}(\widehat {t}_\alpha)$, 
and design optimization will be conducted with respect to minimizing $\mathrm{aVar}(\widehat {t}_\alpha)$. 
This $c$-criterion is commonly used in planning degradation tests when experimenters are interested in accurately estimating 
reliability properties 
of a system over its life cycle. However, it should be noted that the asymptotic variance will depend on $\BS{\beta}$, and thus, 
as a common feature of non-linear models, one is concerned with {\em local} design optimality at some given parameter point $\BS{\beta}$. 
Under a design $\xi$ the asymptotic variance of $\widehat{t}_\alpha$ is given by
\begin{eqnarray}
&&\mathrm{aVar}(\widehat {t}_\alpha)=\BS{c}(\BS{\beta})^{T}\mathbf{M}(\xi,\BS{\beta})^{-1}\BS{c}(\BS{\beta}), \label{c-crit}\\
&&\mbox{where }\ \BS{c}(\BS{\beta})=\frac{\partial t_\alpha}{\partial\BS{\beta}}.\label{c-vector}
\end{eqnarray}
A criterion given by the r.h.s. of (\ref{c-crit}) is called a (local) $c$-criterion. 
Efficient algorithms have been developed to compute a $c$-optimal design, see the numerical example in Subsection \ref{independentoptimaldesign} below.
However, a more explicit formula of the coefficient vector $\BS{c}(\BS{\beta})$ of the criterion has to be provided.
Due to the implicit definition of $t_\alpha$ as the unique solution of $F_T(t_\alpha)=\alpha$, 
the following identity is ensured by the implicit function theorem, 
see \citep{krantz2012implicit}
\begin{equation}
\label{impliciiiiitttt}
\frac{\partial t_\alpha}{\partial \boldsymbol{\beta}}=
\frac{\partial  F_{T}(t_\alpha)}{\partial \boldsymbol{\beta}}\,\big/\,f_T(t_\alpha),\ \mbox{ where }\ f_T(t_\alpha)=
\frac{\partial  F_{T}(t)}{\partial t}\,\Big|_{t=t_\alpha}\,>0.
\end{equation}
From (\ref{distributionnnnnnnfunction}) and (\ref{shape-marginal_l}) one obtains, denoting $Q_1(s,z)=\frac{\partial Q(s,z)}{\partial s}$,
\begin{eqnarray*}     
\frac{\partial  F_T(t_\alpha)}{\partial\beta_{11}} &=&
Q_1\bigl(\gamma_1(x_{u1})t_\alpha,\,z_{10}/\nu_1\bigr)\,Q\bigl(\gamma_2(x_{u2})t_\alpha,\,z_{20}/\nu_2\bigr)\,\gamma_1(x_{u1})t_\alpha,\\
\frac{\partial  F_T(t_\alpha)}{\partial\beta_{21}} &=& x_{u1}\frac{\partial  F_T(t_\alpha)}{\partial\beta_{11}},\\
\frac{\partial  F_T(t_\alpha)}{\partial\beta_{12}}&=&
Q\bigl(\gamma_1(x_{u1})t_\alpha,\,z_{10}/\nu_1\bigr)\,Q_1\bigl(\gamma_2(x_{u2})t_\alpha,\,z_{20}/\nu_2\bigr)\,\gamma_2(x_{u2})t_\alpha,\\
\frac{\partial  F_T(t_\alpha)}{\partial\beta_{22}}&=& x_{u2}\frac{\partial  F_T(t_\alpha)}{\partial\beta_{21}}.
\end{eqnarray*}
Hence, the coefficient vector from (\ref{c-vector}) reads as
\begin{eqnarray}
&&\BS{c}(\BS{\beta})=\bigl(f_T(t_\alpha)\bigr)^{-1}
\bigl(c_1(\BS{\beta})\,(1\,,\,x_{u1})\,c_2(\BS{\beta})\,(1\,,\,x_{u2})\,\bigr)^{T},\ \mbox{ where}\label{c-vector-ex}\\
&&c_l(\BS\beta) = \partial F_T(t_\alpha) / \partial \beta_{1l}\,>0,\,l=1,2. \nonumber
\end{eqnarray}
Together with the block-diagonal structure (\ref{infomatrix-xi}) 
of the information matrices, 
the $c$-criterion from (\ref{c-crit}) becomes 
\begin{equation}
\BS{c}(\BS{\beta})^{T}\mathbf{M}(\xi,\BS{\beta})^{-1}\BS{c}(\BS{\beta})=
\bigl(f_T(t_\alpha)\bigr)^{-2}\sum_{l=1}^2c_l^2\,(1\,,\,x_{ul})\,\textbf{M}_l(\xi_l,\BS{\beta}_l)^{-1}\,(1\,,\,x_{ul})^{T}.
\label{c-crit-ex}
\end{equation}
It follows that a design $\xi^*$ is $c$-optimal w.r.t.~the coefficient vector $\BS{c}(\BS{\beta})$, that is, 
$\xi^*$ minimizes (\ref{c-crit-ex}) over all designs $\xi$ on ${\cal X}=[\,0\,,\,1\,]^2$,
if and only if its marginal designs $\xi_l^*$, $l=1,2$, are $c$-optimal w.r.t. the coefficient vectors $\BS{c}_l=(1\,,\,x_{ul})^{T}$, $l=1,2$,
respectively, that is $\xi_l^*$ minimizes $(1\,,\,x_{ul})\,\textbf{M}_l(\xi_l,\BS{\beta}_l)^{-1}\,(1\,,\,x_{ul})^{T}$ over all designs 
$\xi_l$ on $[0,1]$, $l=1,2$. In particular, $c$-optimality w.r.t.~the coefficient vector $\BS{c}(\BS{\beta})$ 
does not depend on $\alpha$. It should be noted that, under the assumption of independent components, the result can be readily extended to $r>2$ components and to any $s$-out-of-$r$ system, see \citep{shat2021optimal} for further details in this regard. Under the premise that the locally optimal designs $\xi_l^*$ are supported on the endpoints of the design region $[0,1]$, i.\,e., they are of the form $\xi_l^* = \xi_{w_l^*}$, where $\xi_{w_l}$ denotes a design with weight $w_{1l} = w_l$ on $x_{1l} = 0$ and weight $w_{2l} = 1 - w_{1l}$ on $x_{2l} = 1$, \citep {shat2019experimental} stated that the marginal optimal weight $w_l^*$ can be determined analytically by Elfving's theorem \cite{elfving1952},
\begin{equation}
	\label{eq-w-opt-single-marginal}
w_l^* = \frac{(1 + |x_{ul}|) \sqrt{\lambda_l(1,\BS{\beta}_l)}}{(1 + |x_{ul}|) \sqrt{\lambda_l(1,\BS{\beta}_l)} + |x_{ul}| \sqrt{\lambda_l(0,\BS{\beta}_l)}}.
\end{equation}
\subsection{Numerical example}
\label{independentoptimaldesign}
\begin{figure}
	\label{bivariate-independent-gamma-Ft}
	\centering
		\includegraphics[width=0.45\textwidth]{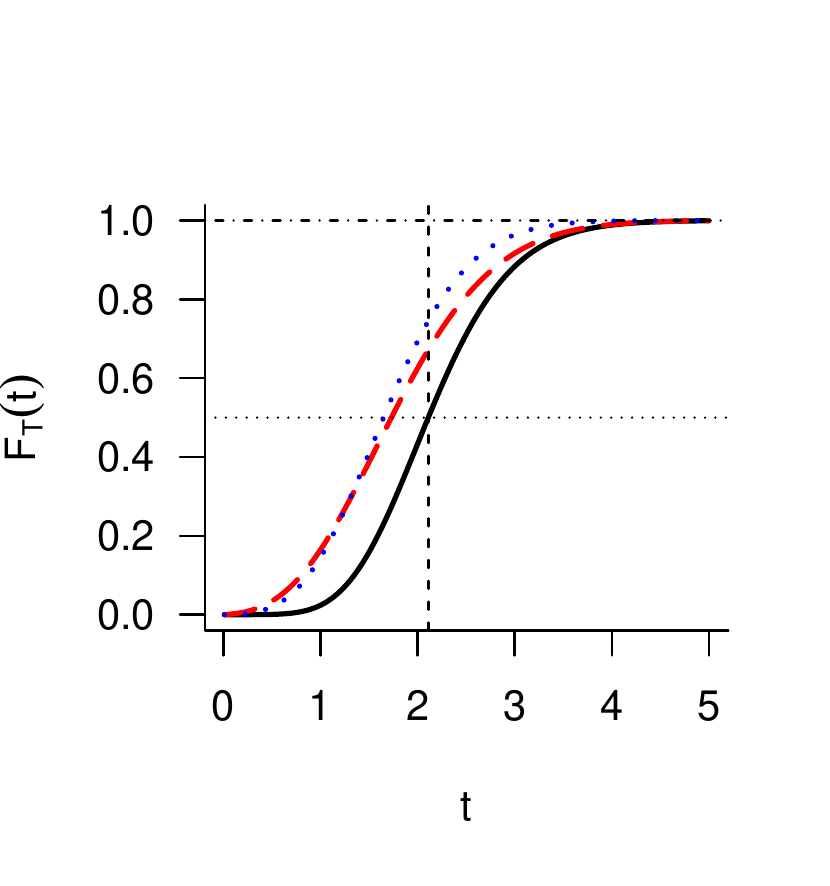}
	\caption{Failure time distribution $F_T(t)$ at the bivariate Gamma model for Example~\ref{independentoptimaldesign}, dashed line: 
                  $F_{T_1}(t)$, dotted line: $F_{T_2}(t)$} 
\end{figure}

The distribution function $F_{T}(t)$ from ({\ref{distributionnnnnnnfunction}}) is plotted for illustration in 
Figure~\ref{bivariate-independent-gamma-Ft} under the nominal values given in Table~\ref{zuuzuzuz123},  the normal use conditions $x_{u1}=-0.60$ and $x_{u2}=-0.50$, and the failure thresholds $z_{10}=4.6$ and $z_{20}=6.25$. 
The median $t_{0.5} = 2.11$ is indicated by a dashed vertical line. Also, the distribution functions $F_{T_l}(t)$ from  
(\ref{eq-failure-time-distribution-margiunal-l}) are shown in the figure. 
We assume that units are observed according to a time plan with $k=4$ time points, and
$t_1=0.02$, $t_2=0.04$, $t_3=0.06$, $t_4=0.1$.
For computing optimal marginal designs $\xi_l^*$ minimizing $(1,x_{ul})\,\mathbf{M}_l(\xi_l,\BS{\beta}_l)^{-1}(1,x_{ul})^{T}$, $l=1,2$,
with nominal values of parameters and constants from  Table \ref{zuuzuzuz123}, the multiplicative algorithm \citep{torsney2009multiplicative} was applied. 
The marginal design interval $[\,0\,,\,1\,]$ was replaced by an equidistant grid with increment equal to $0.05$.  
The obtained optimal marginal designs ${\xi}_1^*$ and ${\xi}_2^*$ are as follows,
\begin{equation}
{\xi}_1^*= \left(\begin{array}{cc}0&1 \\ 0.79&0.21 \end{array} \right)\ \mbox{ and }\ 
{\xi}_2^*= \left(\begin{array}{cc}0&1 \\ 0.91&0.09 \end{array} \right). \label{opt-marg-designs}
\end{equation}
So the locally $c$-optimal designs at $\BS{\beta}$ are given by those designs $\xi^*$ on $\mathcal{{X}}=[\,0\,,\,1\,]^2$
(actually on the product grid of the employed marginal grids) whose marginal designs are equal to $\xi_1^*$ and $\xi_2^*$ from 
(\ref{opt-marg-designs}). One of them is the product design 
\begin{equation}
\label{optimalbivariategamma346543ouztrg}
 {\xi}^* = {\xi}_1^*\otimes{\xi}_2^* 
= \left(\begin{array}{cccc}(0,0)&(0,1)&(1,0)&(1,1) \\ 0.72&0.07&0.19&0.02 \end{array} \right).
\end{equation}
Note that the locally $c$-optimal design is not unique: the set of all designs  with marginal designs given by 
(\ref{opt-marg-designs}) consists of all designs $\xi^*$ supported by the points $(0,0)$, $(0,1)$, $(1.0)$, and $(1,1)$ 
with weights
\[
\xi^*(0,0)=\omega,\ \ \xi^*(0,1)=0.79-\omega,\ \ \xi^*(1,0)=0.91-\omega,\ \ \xi^*(1,1)=\omega -0.70,\quad\mbox{where }\ 0.70\le\omega\le0.79.
\]
For $0.70<\omega<0.79$ the four weights of $\xi^*$ are positive and $\xi^*$ is actually a four-point design. 
The particular value $\omega=0.72$ yields the above product design. The boundary values $\omega=0.70$ and $\omega=0.79$ yield
three-point designs supported by $(0,0)$, $(0,1)$, $(1,0)$ and by  $(0,0)$, $(1,0)$, $(1,1)$, respectively.   

\begin{table}
	\begin{center}
		\caption{Nominal values of the Gamma model with independent marginal components}
		\label{zuuzuzuz123}
		\vspace{3mm}
		\begin{tabular}{|c|c|c||c|c|c|}
			 ${\beta}_{11}$ &$ {\beta}_{12}$&$ {\nu}_{1}$  &$ {\beta}_{21}$ &$ {\beta}_{22}$ &$ {\nu}_{2} $
			\\
			\hline
			 $1.80$ & $1.60$ & $1.24$& $2.80$ & $3.13$ & $1.17$ 
		\end{tabular} 
	\end{center}
\end{table}

\begin{figure}[!tbp]
  \centering
  \begin{minipage}[b]{0.4274632\textwidth}
    \includegraphics[width=\textwidth]{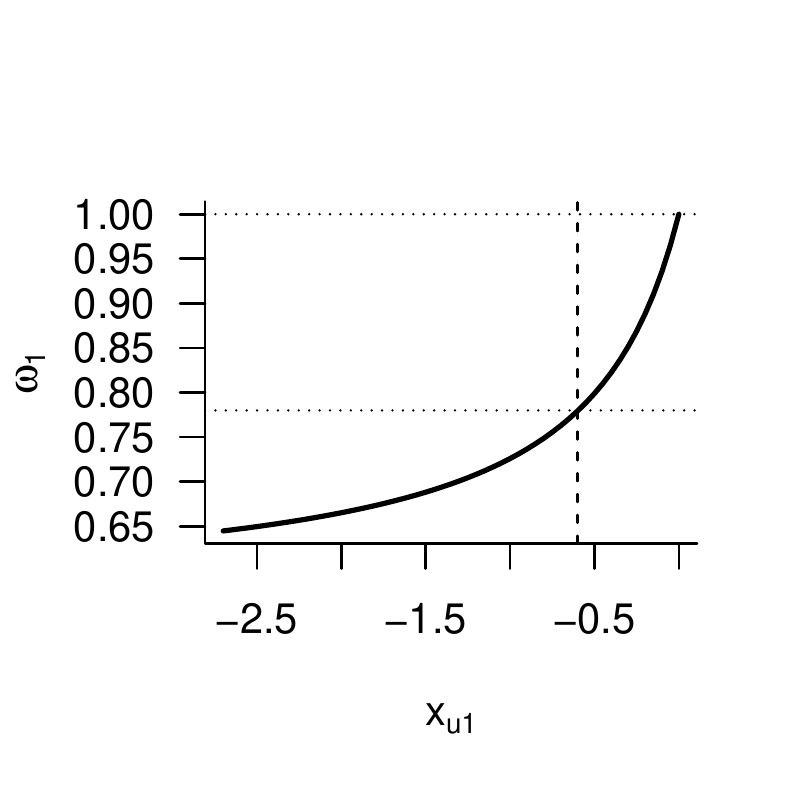}
    \caption{Optimal weights in dependence on $x_{u1}$ for Example~\ref{independentoptimaldesign}} 
\label{fig-weight-x-u1-gammadependent}
  \end{minipage}
  \hfill
  \begin{minipage}[b]{0.4275189\textwidth}
    \includegraphics[width=\textwidth]{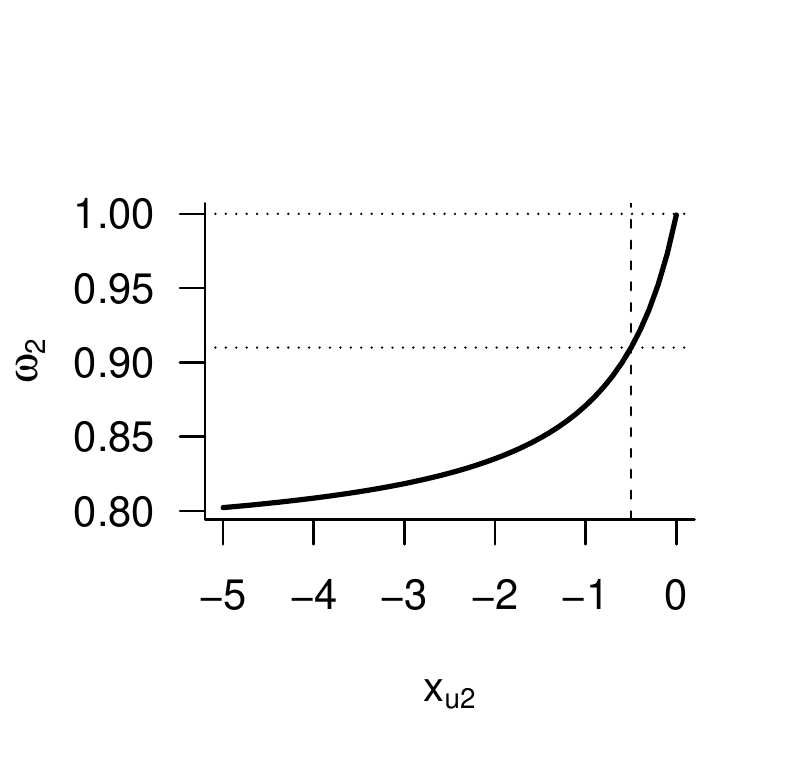}
   \caption{Optimal weights in dependence on $x_{u2}$ for Example~\ref{independentoptimaldesign}}
\label{fig-weight-x-u2-gammadependent}
  \end{minipage}
\end{figure}

When the value of normal use conditions $x_{ul}$, $l=1,2$ are altered within some in intervals of the negative half-line, 
while keeping all other parameters fixed to their nominal values in Table~\ref{zuuzuzuz123},
the optimal marginal designs $\xi_l^*$, $l=1,2$, computed by the algorithm are again supported by the boundary values $0$ and $1$. 
The optimal weight $\omega_1=\xi_1^*(0)$ as a function of $x_{u1}$ is plotted in Figure~\ref{fig-weight-x-u1-gammadependent},
and the optimal weight $\omega_2=\xi_2^*(0)$ as a function of $x_{u2}$ is plotted in  Figure~\ref{fig-weight-x-u2-gammadependent}. 

\begin{figure}[!tbp]
  \centering
  \begin{minipage}[b]{0.4274632\textwidth}
    \includegraphics[width=\textwidth]{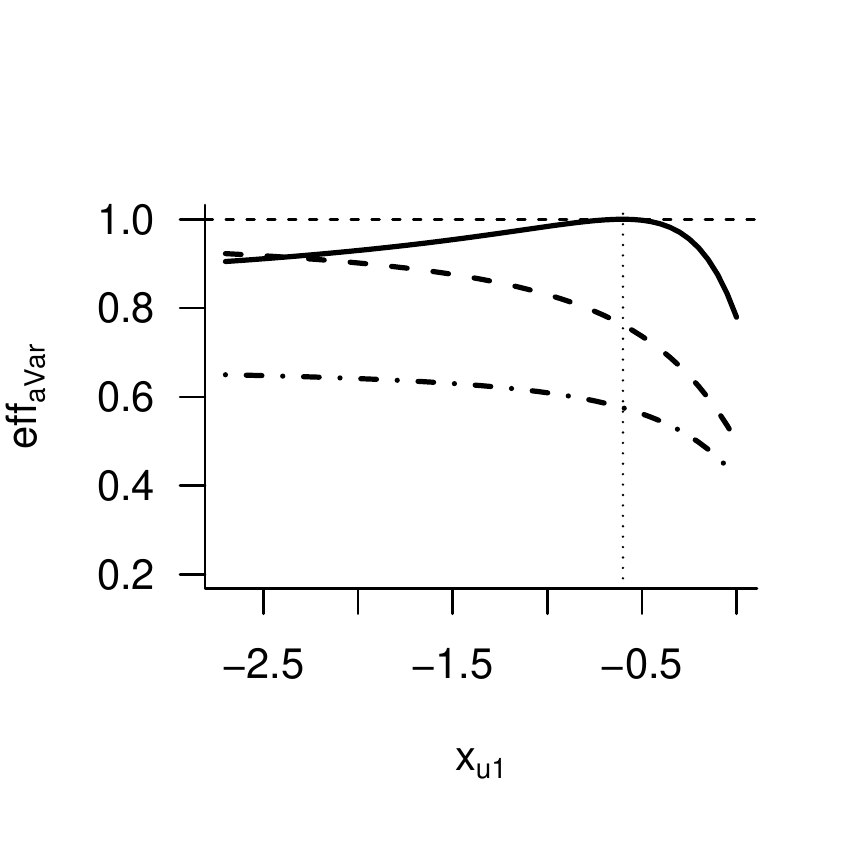}
    \caption{Efficiency of $\xi_1^*$ (solid line), $ \bar\xi_2$ (dashed line) and $ \bar\xi_3$ (dashed and dotted line) in dependence 
             on $x_{u1}$ for Example~\ref{independentoptimaldesign}}
\label{eff-x1-marginalgamma}
  \end{minipage}
  \hfill
  \begin{minipage}[b]{0.42745189\textwidth}
    \includegraphics[width=\textwidth]{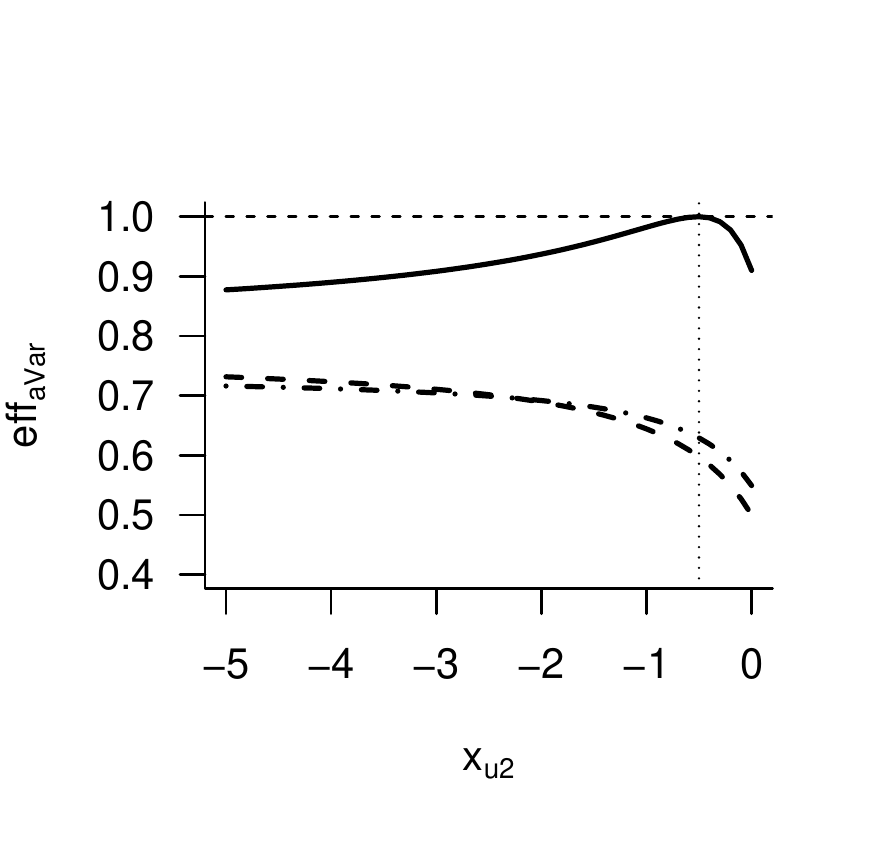}
   \caption{Efficiency of $\xi_2^*$ (solid line), $ \bar\xi_2$ (dashed line) and $ \bar\xi_3$ (dashed and dotted line) in dependence 
              on $x_{u2}$ for Example~\ref{independentoptimaldesign}}
\label{eff-x2-marginalgamma}
  \end{minipage}
\end{figure}

Finally, we examine the influence of varying normal use conditions on the efficiencies of some particular marginal designs $\xi_l$, $l=1,2$.
The efficiency of a marginal design $\xi_l$ at a normal use condition $x_{ul}$, where all other parameters are kept fixed  
according to Table~\ref{zuuzuzuz123}, is defined by
\[
	\mathrm{eff}(\xi_l;  x_{ul}) = \frac{(1,x_{ul})\,\mathbf{M}_l(\xi_l^{(x_{ul})},\BS{\beta}_l)^{-1}(1,x_{ul})^{T}}
{(1,x_{ul})\,\mathbf{M}_l(\xi_l,\BS{\beta}_l)^{-1}(1,x_{ul})^{T}},
\]
where $\xi_l^{(x_{ul})}$ denotes a locally optimal design at $\BS{\beta}_l$, that is, $\xi_l^{(x_{ul})}$ minimizes
$(1,x_{ul})\,\mathbf{M}_l(\widetilde{\xi}_l,\BS{\beta}_l)^{-1}(1,x_{ul})^{T}$ over all marginal designs 
$\widetilde{\xi}_l$ on $[0,1]$, and the present marginal efficiencies may serve as lower bounds for the combined efficiency $\mathrm{eff}(\xi;  \mathbf x_{u})$ of the combined design $\xi$.
In Figure~\ref{eff-x1-marginalgamma} and Figure~\ref{eff-x2-marginalgamma} we plot, respectively, the efficiencies of the 
locally optimal designs 
$\xi_1^*$ and $\xi_2^*$ from (\ref{opt-marg-designs}) (solid line), the efficiencies of the design $\bar\xi_2$ (dashed line) which assigns 
equal weights $1/2$ to the points $0$ and $1$, and the design $\bar\xi_3$ (dashed line) which assigns equal weights $1/3$ to the marginal 
stress levels $0, 0,5$ and $1$. Note that the latter designs $\bar\xi_2$ and $\bar\xi_3$ may serve as standard designs.
The nominal values for $x_{u1}$ and $x_{u2}$ from  Table~\ref{zuuzuzuz123} are indicated in the figures by vertical dotted lines. 
The efficiencies of the optimal designs $\xi_1^*$ and $\xi_2^*$ from (\ref{opt-marg-designs}) 
seem to perform quite well over the ranges of $ x_{u1}$ and $ x_{u2}$, 
respectively. The design $\bar\xi_2$ is preferable for small values of $x_{u1}$ while the design $\bar\xi_3$ performs worse throughout 
for reasonable values of both $x_{u1}$ and $x_{u2}$.

\section{Bivariate Gamma model with dependent components}
\label{dependentcpopulasection345}
Again, let the system under study have two failure modes corresponding to two degradation components,  
but independence of the components will no longer be assumed. 
How to model the case of {\em dependent} degradation components
One would like to have as a model like the following. 
Each marginal degradation component should follow a Gamma process  $Z_{t}^{(l)},\,l=1,2,$ as explained in 
subsection \ref{indepebendebtcopulasectionw5345jhgfdsfsfdsf}.
The joint degradation path of both failure modes 
$\BS{Z}_t=\bigl(Z_{t}^{(1)},Z_{t}^{(2)}\bigr)$ should be a process with independent increments, and the distribution function 
$F^{(h)}(y_1,y_2)$ of an increment $\BS{Z}_{t+h}-\BS{Z}_t$, $t\ge0$, $h>0$, should be given by a fixed copula $C(r,s)$, $0\le r,s\le 1$, 
describing the dependence structure between the marginal processes,
\begin{equation}
F^{(h)}(y_1, y_2) = C\big(F^{(h)}_1(y_1) , F^{(h)}_2(y_2)\big), \ \ y_1,y_2>0,\label{biv-cdf}
\end{equation}
where $F_l^{(h)}(y_l)$, $l=1,2$, denotes the distribution function of the increment $Z_{t+h}^{(l)}-Z_t^{(l)}$ of the marginal Gamma process.
Note that (\ref{biv-cdf}) implies that the bivariate process has stationary increments.  
The reason for using a copula is its ability to provide a flexible and convenient method for combining marginal distributions 
in  a multivariate distribution, see \citep{doi:10.1080/03610918.2010.534227}, see also    \citep{sklar1959fonctions}
for Sklar's Theorem. Two particular copulas are the Frank copula and the Gaussian copula, employed in recent work on degradation modelling, see the corresponding definitions in Subsection \ref{subsec-copula-model} below. However, a copula $C(r,s)$ such that a bivariate process as described exists, is unknown,
unless the independence copula $C(r,s)=rs$ which retrieves the case of independent components. Note that,
by the assumption of independent increments of the bivariate process, the family of bivariate distributions $Q_h$, $h>0$, given by (\ref{biv-cdf}) 
must form a convolution semi-group which, however, is unknown and even not known to exist 
(unless, of course, in case of the independence copula). 
As a way out, we do no longer consider {\em processes} (marginal or bivariate processes),
but restrict to a simple model considering degradations and their increments only at $k$ fixed time points.         
   
\subsection{A simple bivariate copula model}
\label{subsec-copula-model}
Let $k\ge1$ time points be given, $0<t_1<\cdots<t_k$. Denote $\Delta_j=t_j-t_{j-1}$, $j=1,\ldots,k$, where $t_0=0$. Consider
the degradation $Z_j^{(l)}$ at time $t_j$  of the $l$th component and the increments $Y_{jl}= Z_j^{(l)}- Z_{j-1}^{(l)}$,
$j=1,\ldots,k$, $l=1,2$, where $Z_0^{(l)}=0$. For each $l=1,2$, the increments $Y_{1l},\ldots,Y_{kl}$ are independent and Gamma distributed
with parameters as in Section 2. In particular, the shape parameter of the Gamma distribution of $Y_{jl}$ is given by
$\gamma_l(x_l)\,\Delta_j$, where $\mathbf{x}=(x_1,x_2)$ is a normalized bivariate stress variable chosen from the experimental region 
$[\,0\,,\,1\,]^2$, and
\[
\gamma_l(x_l)=\exp\bigl(\beta_{1l}+\beta_{2l}x_l\bigr).
\]
The bivariate increments $\mathbf{Y}_j=(Y_{j1},Y_{j2})$ of the bivariate degradations $\BS{Z}=\bigl(Z_j^{(1)},Z_j^{(2)}\bigr)$, $j=1,\ldots,k$,
are assumed to be independent and follow a distribution according to (\ref{biv-cdf}), that is, the distribution function of $\mathbf{Y}_j$ is given by
\begin{equation}
F_j(y_1,y_2)=C\bigl(F_{j1}(y_1),F_{j2}(y_2)\bigr),\ \ y_1,y_2>0,\label{cdf-biv-a}
\end{equation}     
where $C$ is a given copula and $F_{jl}$ denotes the distribution function of the Gamma distribution with shape parameter 
$\gamma(x_l)\,\Delta_j$ and scale $\nu_l$. 
The copula is assumed to be smooth (sufficiently often continuously differentiable), and thus it has a density
\begin{equation}
c(r,s)= \frac{\partial^2C(r,s)}{\partial r\partial s}, \ \ 0<r,s<1. \label{cop-df}
\end{equation}
Hence it follows that the bivariate increment $\mathbf{Y}_j$ has a density
\begin{equation}
 f_j(\BS{y})= c\bigl( F_{j1}(y_1),F_{j2}(y_2)\bigr)\,f_{j1}(y_1)\,f_{j2}(y_2),\ \ \BS{y}=(y_1,y_2)\in(\,0\,,\,\infty)^2,
\label{biv-df}
\end{equation}
where $f_{jl}$ denotes the Gamma density with shape $\gamma(x_l)\,\Delta_j$ and scale $\nu_l$.

By (\ref{biv-df}) and by independence of the increments, the log-likelihood for the parameter vector 
$\BS{\beta}=(\beta_{11},\beta_{21},\beta_{12},\beta_{22})^{T}$ given the 
values $\BS{y}_1,\ldots,\BS{y}_k$ of the increments $\mathbf{Y}_1,\ldots,\mathbf{Y}_k$ and under the stress condition 
$\mathbf{x}=(x_1,x_2)\in[\,0\,,\,1\,]^2$, reads as
\begin{equation}
\ell\bigl(\BS{\beta};\BS{y}_1,\ldots,\BS{y}_k,\,\mathbf{x}\bigr) =   
\sum_{j=1}^{k} \bigg[\exp \Big(c\big(F_{j1}(y_{j1}), F_{j2}(y_{j2})\big)\Big) +\sum_{l=1}^2\exp\Big(f_{jl}(y_{jl})\Big)\biggr].
\label{log-like-li-hooood}
\end{equation}
The following definitions present two particular copulas (in two dimensions) to be considered in further applications: the Frank copula and the Gaussian copula. 

\begin{definition}
\label{Frank-cop}
The Frank copula, which is a very common Archimedean copula for bivariate data, is utilized to describe the dependence relation 
between marginal failure modes. The bivariate Frank copula is defined as
\begin{equation}
\label{cdf-copula-Frank}
C(r,s)= -\frac{1}{\varkappa}\exp \left(1+ \frac{\big(e^{-\varkappa r}-1\big)\big(e^{-\varkappa s}-1\big)}{e^{-\varkappa}-1}\right)
\end{equation}
where $\varkappa \in (-\infty, \infty) \backslash  \{0\}$ is a fixed copula dependence parameter. The density from 
(\ref{cop-df}) becomes
\begin{equation} 
\label{density-copula-Frank}
c\big(r,s\big)=
\frac{\varkappa \big(1-e^{-\varkappa }\big) e^{ -\varkappa (r+s)}}{\Big(1-e^{-\varkappa}-
\big(1-e^{ -\varkappa r}\big)\big(1-e^{ -\varkappa s}\big)\Big)^2}
\end{equation}
\end{definition}

\begin{definition}
\label{Gaussian-cop}
The Gaussian copula employs a correlation parameter $\rho$ defining a  positive definite correlation matrix 
\[
\BS{\Sigma}=\begin{bmatrix} 
 1 &\rho\\
\rho &1
\end{bmatrix},\,\,\,\,\,-1<\rho<1.
\]
Denote by $\Phi$ the standard normal distribution function, and denote by $F_{\BS{0},\BS{\Sigma}}$ the distribution function of the bivariate normal
distribution with expectation $\BS{0}$ and covariance matrix $\BS{\Sigma}$, that is,
\begin{equation}
\label{copula of gaussian2222}
F_{\BS{0},\BS{\Sigma}}(a,b)=(2\pi)^{-1}(\det(\Sigma))^{-1/2}\int_{-\infty}^a\int_{-\infty}^b \exp\big(-{\textstyle\frac{1}{2}} 
\BS{z}^T \Sigma^{-1} \BS{z} \big)\,d\BS{z}.
\end{equation}
Then, the Gaussian copula reads as
\begin{equation}
\label{copula of gaussian}
C(r,s)=F_{\BS{0},\BS{\Sigma}}\big(\Phi^{-1}(r),\Phi^{-1}(s)\big),\,\,\,\,\,r,s\in(0,1).
\end{equation}
Its density according to (\ref{cop-df}) is given by
\begin{equation}
\label{density-copula of gaussian}
c(r,s)=
\frac{(2\pi)^{-1}(\det(\Sigma))^{-1/2}\exp\big(-\frac{1}{2}   \bigl(\Phi^{-1}(r),\Phi^{-1}(s)\bigr) \Sigma^{-1} 
\bigl(\Phi^{-1}(r),\Phi^{-1}(s)\bigr)^{T} \big)}{ \phi\bigl(\Phi^{-1}(r)\bigr)\,
\phi\bigl(\Phi^{-1}(s)\bigr)},
\end{equation}
where $\phi$ denotes the standard normal density.  
The normal copula space provides a flexible and convenient method for combining marginal distributions in  a 
multivariate distribution, see \citep{doi:10.1080/03610918.2010.534227}. Using the Gaussian copula in our 
bivariate Gamma model, the resulting  density (\ref{biv-df}) of a bivariate increment was employed   
in \citep{adegbola2019multivariate}.
\end{definition}

\subsection{Information matrix}
\label{information-copulaaaaa}
From the log-likelihood (\ref{log-like-li-hooood}) we calculate  
the elemental Fisher information matrix of $\mathbf{x}$ at $\BS{\beta}$,
\begin{equation}
\mathbf{M}(\mathbf{x},\BS{\beta}) = -{\rm E}\Bigl(\frac{\partial^2\ell(\BS{\beta}; \BS{Y}_1,\ldots,\BS{Y}_k,\,\mathbf{x})}{
\partial\BS{\beta}\partial\BS{\beta}^{T}}\Bigr), \label{elem-info-a}
\end{equation}
The symbol $\mathbf{M}_{\rm\scriptsize ind}(\mathbf{x},\BS{\beta})$
will used for the elemental information matrix from the model with independent components studied in Section 2. 
In fact, on the r.h.s. of (\ref{log-like-li-hooood}), the second term (double sum over $j=1,\ldots,k$ and $l=1,2$) yields, 
after (twice) partial differentiation, taking the expectation and putting a minus sign in front,
the information matrix from (\ref{elem-info-ex_a}) since the expectation of that term depend only on the marginal 
distributions of $\mathbf{Y}_j$, $j=1,\ldots,k$, which are the same Gamma distributions as in Section 2.  
It remains to calculate the matrix 
\begin{equation}
{\rm E}\Bigl(\frac{\partial^2 \exp c\bigl((F_{j1}(Y_{j1}),F_{j2}(Y_{j2})\bigr)}{\partial\BS{\beta}\,\partial\BS{\beta}^{T}}\Bigr).
\label{info-matrix-task}
\end{equation}
Here  $F_{jl}$ denotes the distribution function of the Gamma distribution with shape $\gamma_l(x_l)\,\Delta_j$ and scale $\nu_l$, 
$f_{jl}$ denotes its density, and $\gamma_l(x_l)=\exp(\beta_{1l}+\beta_{2l}x_l\bigr)$. 
Formulas for (\ref{info-matrix-task}) are derived in \ref{part444442343},
which involve two-dimensional integrals. From this, the information matrix (\ref{elem-info-a}) reads as
\begin{eqnarray}
&&\mathbf{M}(\mathbf{x},\BS{\beta}) = \mathbf{H}(\mathbf{x},\BS{\beta}) + \mathbf{M}_{\rm\scriptsize ind}(\mathbf{x},\BS{\beta}),
\label{elem-info-b}\\
&&\mbox{where }\ \mathbf{H}(\mathbf{x},\BS{\beta}) = 
\left[\begin{array}{ll}\mathbf{H}_1(\mathbf{x},\BS{\beta}) & \mathbf{H}_{12}(\mathbf{x},\BS{\beta})\\
\mathbf{H}_{12}^{T}(\mathbf{x},\BS{\beta}) & \mathbf{H}_2(\mathbf{x},\BS{\beta})\end{array}\right],\nonumber\\
&&\mathbf{H}_l(\mathbf{x},\BS{\beta})=\varphi_l(\mathbf{x},\BS{\beta})\,\bigl(1, x_l\bigr)^{T}\bigl(1, x_l\bigr),
\ l=1,2,\ \ 
\ \ \mathbf{H}_{12}(\mathbf{x},\BS{\beta}) = \varphi_{12}(\mathbf{x},\BS{\beta})\,\bigl(1, x_1\bigr)^{T}\bigl(1, x_2\bigr),
\nonumber\\
&&\varphi_l(\mathbf{x},\BS{\beta})=\gamma_l^2(x_l)\sum_{j=1}^k\int_0^\infty\int_0^\infty 
\frac{c_l^2\bigl(F_{j1}(y_1),F_{j2}(y_2)\bigr)}{c\bigl(F_{j1}(y_1),F_{j2}(y_2)\bigr)}\,
\Bigl(\frac{\partial F_{j1}(y_1)}{\partial \gamma_l}\Bigr)^2\,f_{j1}(y_1)\,f_{j2}(y_2)\,{\rm d}y_1\,{\rm d}y_2,
\ \ l=1,2,\nonumber
\end{eqnarray}
and
\begin{eqnarray*}
&&\varphi_{12}(\mathbf{x},\BS{\beta})=\\
&&\gamma_1(x_1)\,\gamma_2(x_2) \sum_{j=1}^k\int_0^\infty\int_0^\infty
\Bigl[\frac{c_1\bigl(F_{j1}(y_1),F_{j2}(y_2)\bigr)\,c_2\bigl(F_{j1}(y_1),F_{j2}(y_2)\bigr)}
{c\bigl(F_{j1}(y_1),F_{j2}(y_2)\bigr)}\,\frac{\partial F_{j1}(y_1)}{\partial\gamma_1}\,\frac{\partial F_{j2}(y_2)}{\partial\gamma_2}
\,f_{j1}(y_1)\,f_{j2}(y_2)\\
&&\phantom{xxxxxxxxxxxxxxxxxxxxxxx}-\,c\bigl((F_{j1}(y_1),F_{j2}(y_2)\bigr)\,\frac{\partial f_{j1}(y_1)}{\partial\gamma_1}\,
\frac{\partial f_{j2}(y_2)}{\partial \gamma_2}\Bigr]\,{\rm d}y_1\,{\rm d}y_2.
\end{eqnarray*}
such that $c_1(r,s)$ and $c_2(r,s)$ denote the first order partial derivatives of the copula density $c(r,s)$, that is,
\[
c_1(r,s)=\frac{\partial c(r,s)}{\partial r}\ \mbox{ and }\ c_2(r,s)=\frac{\partial c(r,s)}{\partial s}, \ \ \ 0<r,s<1.
\]
Formulas for the partial derivatives $\partial F_{jl}(y_1)\big/\partial \gamma_l$ and $\partial f_{jl}(y_1)\big/\partial \gamma_l$
are given in \ref{part444442343}. Note that in case of equidistant time points $t_1,\ldots,t_k$, that is, $\Delta_j=\Delta$ for $j=1,\ldots,k$,
the distribution functions and densities $F_{jl}$ and $f_{jl}$, respectively, are independent of $j$, and the above formulas simplify  
in that case.  

As usual, if $\xi$ is an (approximate) design on $[\,0\,,\,1\,]^2$ with support points $\mathbf{x}_1,\ldots,\mathbf{x}_m$
and corresponding weights $w_i$, $i=1,\ldots,m$, the information matrix of $\xi$ at a parameter point $\BS{\beta}$
is given by
\begin{equation}
\mathbf{M}(\xi,\BS{\beta})=\sum_{i=1}^mw_i\,\mathbf{M}(\mathbf{x}_i,\BS{\beta}).\label{design-info-matrix}
\end{equation}
In contrast to the settings of independent response components in Section \ref{indepebendebtcgpulasectionw5345jhg}, the $D$-optimality criterion will be applied, instead of the $c$-criterion, for the current settings of Copula-based bivariate degradation models. The main reason behind that is the difficulty to accurately define the continuous failure time variable $T$, and, hence, the quantile $t_\alpha$, under the assumptions of dependent marginal failure modes based on Copula functions. Accordingly, we are adopting the $D$-criterion for the numerical calculations in Example~\ref{Frank-cop-D-opt} and Example~\ref{Gaussian-cop-D-opt}.

\subsection{Local D-optimality}
\label{optimalitycriterionnnnnff456gdfg}
For a given parameter point $\BS{\beta}$, a design $\xi^*$ is called locally $D$-optimal at $\BS{\beta}$ if 
$\xi^*$ maximizes $\det\bigl(\mathbf{M}(\xi,\BS{\beta})\bigr)$ over all designs $\xi$. 
For numerical computation of a locally $D$-optimal design we used the multiplicative algorithm, where the design region
$[\,0\,,\,1\,]^2$ is discretized by a grid with $0.05$ increments in both dimensions. The elemental information matrices from
(\ref{elem-info-b}) were computed by numerical integration in two dimensions. We employed the Frank copula and the 
Gaussian copula from based on \ref{Frank-cop} and \ref{Gaussian-cop}, respectively. 

\begin{example}
\label{Frank-cop-D-opt} 
Let $C(r,s)$ be the Frank copula from (\ref{cdf-copula-Frank}). Its density $c(r,s)$ is given by
(\ref{density-copula-Frank}). By straightforward calculations, one obtains the first order partial derivatives  
$c_1(r,s)=\partial c(r,s)\big/\partial r$ and $c_2(r,s)=\partial c(r,s)\big/\partial s$, 
\begin{equation}
c_1(r,s)=
\frac{\varkappa^2\big(1-e^{-\varkappa}\big)e^{-\varkappa(r+s)}\big[(1+e^{-\varkappa r})(1-e^{-\varkappa s})-(1-e^{-\varkappa})\big]}
{\big[1-e^{-\varkappa}-\,\big(1-e^{-\varkappa r}\big)\big(1-e^{-\varkappa s}\big)\big]^3},\quad c_2(r,s)=c_1(s,r).
\end{equation}

\begin{table}
	\begin{center}
		\caption{Nominal values of the bivariate Gamma model with Copula function}
		\label{zuuzuzuz456546}
		\vspace{3mm}
		\begin{tabular}{|c|c|c||c|c|c||c|c||c|c|}
			 ${\beta}_{11}$ &$ {\beta}_{12}$&$ {\nu}_{1}$  &$ {\beta}_{21}$ &$ {\beta}_{22}$ &$ {\nu}_{2} $ &$ {\varkappa} $&$ {\rho} $
			\\
			\hline
			 $0.30$ & $0.90$ & $1.17$& $0.80$ & $0.10$ & $1.15$ &  $-0.40$ &  $-0.10$
		\end{tabular} 
	\end{center}
\end{table}

Choosing $k=4$ equidistant time points $t_1=0.05$, $t_2=0.10$, $t_3=0.15$, $t_4=0.20$, 
and the nominal values of the parameter vector $\boldsymbol{\beta}$ in Table~\ref{zuuzuzuz456546},
numerical computations with the multiplicative algorithm were done for local D-optimal design. 
The obtained locally D-optimal design is a uniformly weighted 6-point design,
\begin{equation}
\label{optimal-cont.-frank}
  {\xi}_{_D}^* = \left(\begin{array}{cccccc}(0,0)&(0,1)&(0.5,0)&(0.5,1)&(1,0)&(1,1) \\ 0.166&0.166&0.166&0.166&0.166&0.166\end{array} \right)
\end{equation}

\end{example}

\begin{example}
\label{Gaussian-cop-D-opt}
Let $C(r,s)$ be the Gaussian copula from (\ref{copula of gaussian}) with parameter value $\rho=-0.1$.
Its density is given by (\ref{density-copula of gaussian}), and the first order partial derivatives of the latter are given by
\[
c_1(r,s)=\frac{\rho}{1-\rho^2}\,c(r,s)\,\frac{\Phi^{-1}(s)-\rho\Phi^{-1}(r)}{\phi\bigl(\Phi^{-1}(r)\bigr)},
\quad c_2(r,s)=c_1(s,r).
\]
As is the preceeding example, we choose  $k=4$ equidistant time points $t_1=0.05$, $t_2=0.10$, $t_3=0.15$, $t_4=0.20$, 
and the nominal values of the parameter vector $\boldsymbol{\beta}$ from  Table~\ref{zuuzuzuz456546}.
The locally D-optimal design obtained with the multiplicative algorithm has the same six support points as 
that for Example~\ref{Frank-cop-D-opt}, with non-uniform weights, as
\begin{equation}
\label{optimal-cont.-gaussian}
  {\xi}_{_D}^* = \left(\begin{array}{cccccc}(0,0)&(0,1)&(0.5,0)&(0.5,1)&(1,0)&(1,1) \\ 0.20&0.20&0.16&0.16&0.18&0.09\end{array} \right)
\end{equation}

\end{example}
Due to the difficulty of accurately deriving the information matrix \ref{elem-info-b} for the Copula-based models \ref{Frank-cop} and \ref{Gaussian-cop} with multiple observations, we consider in Section~\ref{bivariatewithcorrelation} a simplified approach with binary outcomes which facilitates the derivations of the corresponding information matrix and, hence, considerably reduce the calculations time.

\section{Copula-based gamma model with binary outcomes}
\label{bivariatewithcorrelation}
\subsection{Model formulation}
\label{bivariatewithcorrelation-introduction}
In this section, we consider the model from Section \ref{dependentcpopulasection345}, but now  
the measurements of bivariate degradations $\BS{Z}_j =(Z_j^{(1)},Z_j^{(2)}),\,j=1,...,k,$
are reduced to the information on whether or not the marginal degradation paths have reached or exceeded given
thresholds ${z_{10}} > 0$ and ${z_{20}} > 0$, respectively, at each time $t_j,\,j=1,...,k$. 
This information is equivalently reflected by two discrete variables $U$ and $V$ with values in $\{1,...,k,k+1\}$,
where $U$ (resp. $V$) gives the first time label $j$ such that the marginal degradation $Z_j^{(1)}$ (resp. $Z_j^{(2)}$) 
has reached or exceeded the threshold $z_{01}$ (resp. $z_{02}$), and  
the value $k + 1$ expresses that failure did not occur until time $t_k$. That is, we define
\begin{eqnarray*}
U &=& \min\Big\{j\in\{ 1,...,k\}:  Z_j^{(1)}\geq z_{10}\Big\},\\
V &=& \min\Big\{j\in\{ 1,...,k\}: Z_j^{(2)}\geq {z_{20}}\Big\},
\end{eqnarray*}
where the minimum of the empty set is defined to be $k+1$. The joint distribution of $U, V$ is given
by the probabilities $P_{u,v} = \Pr(U=u,V=v),\, u,v\in\{1,...,k,k+1\}$. Below we will see that their calculation involves 
multi-dimensional integrals over polyhedral regions which are difficult to handle theoretically as well as numerically.
A slight simplification of the integration regions is gained by considering the probabilities  
$$Q_{u,v} = \Pr(U\leq u,V\leq v)\quad \mbox{for}\quad 1\leq u,v\leq k+1.$$    
Note that $Q_{k+1,v}=\Pr(V\leq v)$ and $Q_{u,k+1}=\Pr(U\leq u)$, and especially $Q_{k+1,k+1}=1$. The probabilities $P_{u,v}$ 
are obtained from the  $Q_{u,v}$ by
\begin{equation}\label{probability-w.r.t.-Q}\\P_{u,v}=Q_{u,v}-Q_{u,v-1}-Q_{u-1,v}+Q_{u-1,v-1}\,\, \mbox{ for}\,\, 1\leq u,v\leq k+1\end{equation}
where $Q_{0,0}=Q_{0,v}=Q_{u,0}=0
$ for $1\leq u,v\leq k+1$. By the two equivalences, for any $u,v\in\{1,...,k\}$,
\[
U\leq u \ \Longleftrightarrow \ Z_u^{(1)}\geq {z_{10}},\quad
V\leq v \  \Longleftrightarrow \  Z_v^{(2)}\geq {z_{20}},
\]
and writing the degradations as sums of increments, $Z_u^{(1)}=\sum_{j=1}^u Y_{j1}$ and $Z_v^{(2)}=\sum_{j=1}^v Y_{j2}$,
we get for all $u,v \in\{1,...,k\},$

\begin{equation}
Q_{u,v}=\int_{A_{u,v}}\prod_{j=1}^k f_j(\mathbf{Y}_j)\ d {\BS y}{_{1}}\cdots{\BS y}{_{k}},\label{QQQequation}
\end{equation}
where 
\[  
A_{u,v}=\Big\{({\BS y}{_{1}},...,{\BS y}{_{k}})\in(0,\infty)^{2k}: 
\sum_{j=1}^u y_{j1}\geq {z_{10}}, \sum_{j=1}^v y_{j2}\geq {z_{20}} \Big \},
\]
and $f_j$ denotes the density of the bivariate increment $\mathbf{Y}_j=(Y_{j1},Y_{j2})$ from  (\ref{biv-df}).
For $u=k+1$ or $v=k+1$, a calculation of $Q_{k+1,v}$ or $Q_{u,k+1}$ involves only the marginal degradations, which are Gamma distributed, 
\begin{eqnarray*}
Q_{k+1,v} &=& {\rm Pr}\bigl(Z_v^{(2)}\ge z_{20}\bigr)= 
\frac{\Gamma\bigl(\gamma_2(x_2)\, t_v , y_2/\nu_2)}   {\Gamma\bigl(\gamma_2(x_2)\, t_v},\quad 1\leq v\leq k,\\
Q_{u,k+1} &=& \frac  {\Gamma\bigl(\gamma_1(x_1)\, t_u , y_1/\nu_1)}   {\Gamma\bigl(\gamma_1(x_1)\, t_u)},\quad 1\leq u\leq k.
\end{eqnarray*}

\subsection{Information matrix}
\label{info-binary-stand}
The log likelihood of the bivariate discrete variable $(U,V)$ is given by\\
\begin{equation}
\ell(\BS{\beta};u,v,\mathbf{x})=
\exp P_{u,v}(\mathbf{x},\BS{\beta}) 
 \end{equation}\\
where now we observe the dependence of the probabilities $P_{u,v}, 1\leq u,v\leq k+1,$ on the design variable 
$\mathbf{x}=(x_1,x_2)$ and the parameter vector $\BS{\beta}=(\beta_{11},\beta_{12},\beta_{21},\beta_{22})^T$. 
The elemental information matrix of $\mathbf{x}$ at a parameter point $\BS{\beta}$ is given by
\begin{equation}
\mathbf{M}(\mathbf{x},\BS{\beta})=
\mbox{E}\Bigg[\Bigg(\frac{\partial\ell(\BS{\beta};u,v,\BS x)}{\partial \BS\beta}\Bigg)
\Bigg(\frac{\partial\ell(\BS{\beta};u,v,\BS x)}{\partial \BS\beta}\Bigg)^T\Bigg].
\end{equation}
We can decompose $\ell$, as a function of $\BS \beta$, according to
\[
\BS{\beta}\ \longrightarrow\ \BS{\gamma}=(\gamma_1,\gamma_2)^T\ \longrightarrow\ 
\BS{P}=(P_{11},...,P_{uv},...,P_{k+1,k+1})^T\ \longrightarrow\ \ell,
\]
where the $P_{uv}$, $1\le u,v\le k+1$, have been arranged in lexicographic order, say, to form the vector $\BS{P}$. 
By the chain rule a factorization of the gradient $\partial\ell(\BS{\beta};u,v,\BS x)/\partial \BS\beta$ results, 
\[
\frac{\partial\ell(\BS{\beta};u,v,\BS x)}{\partial \BS\beta}=\BS A \BS B \BS C,
\]
with matrices $\BS{A}$, $\BS{B}$ and a column vector $\BS{C}$,
\begin{eqnarray*}
\BS A &=&  \BS A (\mathbf{x},\BS{\beta}) =\frac{\partial \BS\gamma}{\partial \BS \beta}= 
\begin{bmatrix} 
 \gamma_1 &0\\
x_1\gamma_1&0\\
0 &\gamma_2\\
0&x_2\gamma_2\\
\end{bmatrix} ,\\
\BS{B} &=& \BS{B} (\mathbf{x},\BS{\beta}) =\frac{\partial \BS{P}}{\partial \BS{\gamma}}= 
\begin{bmatrix} 
 \frac{\partial P_{uv}}{\partial \gamma_1}(1\leq u,v\leq k+1)\\
 \frac{\partial P_{uv}}{\partial \gamma_2}(1\leq u,v\leq k+1)\\
\end{bmatrix} ,\\
\BS C &=& C (\mathbf{x},\BS{\beta},u,v) =\frac{\partial \ell}{\partial \BS P_{(1\le u,v\le k+1)}}=\Big(
\frac{1}{P_{uv}}\Big)^{T}.
\end{eqnarray*}
Note that the two rows of $\BS{B}$ and the column vector $\BS{C}$ have components indexed by the pairs $(u,v)$
arranged in lexicographic order. It follows that
\begin{equation}
\mathbf{M}(\mathbf{x},\BS{\beta})=\BS A \BS B\,\mbox{E}(\BS C \BS C^T)\,\BS B^T \BS A^T,
\end{equation} 
and 
\[
\mbox{E}(\BS C \BS C^T)=\mbox{diag}\Big(\frac{1}{P_{uv}}(1\leq u,v\leq k+1)\Big).
\]
Again, for a design $\xi$ with support points $\mathbf{x}_i$ and weights $w_i$, $i=1,\ldots,m$,
the information matrix of $\xi$ at $\BS{\beta}$ is given by
\[
\mathbf{M}(\xi,\BS{\beta})= \sum_{i=1}^mw_i\,
\mathbf{M}(\mathbf{x}_i,\BS{\beta}).
\]
In order to obtain explicit formulas for the entries of $\BS{B}(\mathbf{x},\BS{\beta})$, that is, 
the  partial derivatives $\partial P_{uv}\big/\partial\gamma_l$,
we consider the corresponding partial derivatives of the probabilities $Q_{u,v}$ from (\ref{QQQequation}). One gets
\begin{eqnarray}
\frac{\partial Q_{u,v}}{\partial \gamma_l} &=& 
\int_{A_{u,v}}\frac{\partial}{\partial \gamma_l}\prod_{j=1}^k f_j(\mathbf{Y}_j)\,{\rm d}\BS{y}_1...{\rm d}\BS{y}_k\nonumber\\
&=& \int_{A_{u,v}}\sum_{i=1}^{k}\Big[\prod_{j\neq i}f_j(\mathbf{Y}_j)\Big]\frac{\partial f_i(\BS{y}_i)}{\partial \gamma_l}\ 
{\rm d}\BS{y}_1...\BS{y}_k,\label{derv-q-u,v}
\end{eqnarray}
and by (\ref{biv-df}), 
\begin{equation}
\label{derv-q-u,v-a}
\begin{split}
&\frac{\partial f_j(\mathbf{Y}_j)}{\partial \gamma_1} = \left[c_1\bigl(F_{j1}(y_{j1}),F_{j2}(y_{j2})\bigr)\,
\frac{\partial F_{j1}(y_{j1})}{\partial\gamma_1}\,f_{j1}(y_{j1})
\,+\,c\bigl(F_{j1}(y_{j1}),F_{j2}(y_{j2})\bigr)\,\frac{\partial f_{j1}(y_{j1})}{\partial\gamma_1}\right]\,f_{j2}(y_{y2}),\\
&\frac{\partial f_j(\mathbf{Y}_j)}{\partial \gamma_2} = \left[c_2\bigl(F_{j1}(y_{j1}),F_{j2}(y_{j2})\bigr)\,
\frac{\partial F_{j2}(y_{j2})}{\partial\gamma_2}\,f_{j2}(y_{j2})
\,+\,c\bigl(F_{j1}(y_{j1}),F_{j2}(y_{j2})\bigr)\,\frac{\partial f_{j2}(y_{j2})}{\partial\gamma_2}\right]\,f_{j1}(y_{j1}).
\end{split}
\end{equation}\\
However, due to the $2k$-dimensional integration in (\ref{derv-q-u,v}) 
the calculation of information matrices is not tractable when $k>1$. 
Therefore, we consider now the simple case $k=1$ of a single measurement. Then, we have one bivariate increment 
$\BS{Y} = (Y_1, Y_2)$, and the distribution function of $\BS{Y}$ is given by
\[
C(F_1(y_1); F_2(y_2)),\,\,\,\,\,\, y_1, y_2 \in ( 0 , \infty).
\]
The probabilities $P_{uv}$, $u,v\in\{1,2\}$, can be expressed by the latter joint distribution function and the marginal distribution 
functions $F_1$ and $F_2$,
\begin{equation} 
\begin{split} 
\label{p_uv}
P_{_{2,2}}=\mathrm{P}\big(Y_1<{z_{10}}, Y_2<{z_{20}}\big)=&C\big(F_1({z_{10}}),F_2({z_{20}})\big),\\
P_{_{1,2}}=\mathrm{P}\big(Y_1\geq{z_{10}}, Y_2<{z_{20}}\big)=&F_2({z_{20}})-C\big(F_1({z_{10}}),F_2({z_{20}})\big),\\
P_{_{2,1}}=\mathrm{P}\big(Y_1<{z_{10}}, Y_2\geq{z_{20}}\big)=&F_1({z_{10}})-C\big(F_1({z_{10}}),F_2({z_{20}})\big),\\
P_{_{1,1}}=\mathrm{P}\big(Y_1\geq{z_{10}}, Y_2\geq{z_{20}}\big)=&1-F_1({z_{10}})-F_2({z_{20}})+C\big(F_1({z_{10}}),F_2({z_{20}})\big).
\end{split} 
\end{equation} 
The partial derivatives ${\partial P_{u,v}}/{\partial\alpha_l}$ are easily obtained from the partial derivatives 
$\partial F_l(z_{l0})/\partial\alpha_l$ and the partial derivatives of the copula, $C_1(r,s)=\partial C(r,s)/\partial r$
and $C_2(r,s)=\partial C(r,s)/\partial s$, since by the chain rule  
\begin{equation}
\frac{\partial}{\partial \gamma_l} C\big(F_1({z_{10}}),F_2({z_{20}})\big) = C_l\big(F_1(z_{10}),F_2(z_{20})\big)\, 
\frac{\partial F_l(z_{l0})}{\partial\gamma_l},\ \ l=1,2.\label{p_uv-dev}
\end{equation} 
In particular, when $C$ is the Frank copula with parameter $\varkappa$, then by straightforward calculation,
\begin{equation}
C_1(r,s) = \frac{e^{-\varkappa r}\,(  e^{-\varkappa s}-1)}{  e^{-\varkappa}-1  + (  e^{-\varkappa r}-1)\,( e^{-\varkappa s}-1)}
\ \mbox{ and }\ C_2(r,s) = C_1(s,r).\label{frank-copulaaa-rewritten-derv}
\end{equation}
When $C$ is the Gaussian copula with correlation parameter $\rho$, then one obtains (see \ref{part444442343})
\begin{equation}
C_1(r,s) = \Phi\Bigg(\frac{\Phi^{-1}(s)-\rho\Phi^{-1}(r)}{\sqrt{1-\rho^2}}\Bigg)\,\,\,\,\,\mbox{and}\,\,\,\,\,
C_2(r,s) = C_1(s,r).
\label{Gaussian-cop-C_l}
\end{equation}

\subsection{Local D- or $c$-optimal designs when $k=1$}
For our simple binary model ($k=1$) employing the Frank copula or the Gaussian copula, locally $D$- or $c$-optimal designs
are presented in the example below. A locally $D$-optimal design $\xi^*_D$ at a given parameter point $\BS{\beta}$ maximizes
$\det\bigl(\mathbf{M}(\xi,\BS{\beta})\bigr)$ over all designs $\xi$. A locally $c$-optimal design $\xi^*_c$ at
$\BS{\beta}$ minimizes $\BS{c}^{T} \mathbf{M}(\xi,\BS{\beta})^{-1}\BS{c}$ over all designs $\xi$,
where $\BS{c}$ is a given nonzero column vector of dimension four. Here the coefficient vector $\BS{c}$ is chosen such that
the $c$-criterion represents the asymptotic variance of the maximum likelihood estimator $\widehat{P}_{11}$ of the joint failure probability
$P_{11}=P_{11}(\mathbf{x}_u,\BS{\beta})$ at normal use conditions $\mathbf{x}_u=(x_{u1},x_{u2})$. That is,
\[
\BS{c}=\frac{\partial P_{11}(\mathbf{x}_u,\BS{\beta})}{\partial\BS{\beta}}=\bigl(c_1.(1,x_{u1})\,,\,c_2.(1,x_{u2})\bigr)^{T},
\]
where 
\[
c_l=\gamma_l(x_{ul})\,\Delta_1\frac{\partial P_{11}}{\partial\gamma_l},\ \ l=1,2.
\]
The partial derivatives $\partial P_{11}/\partial\gamma_l$ can be evaluated using formulas (\ref{p_uv}), (\ref{p_uv-dev}), 
and (\ref{eqA-5}).
\begin{example}
\label{sec:3.4656}
For obtaining numerically optimal designs, the multiplicative algorithm with an equidistant  grid of 0.05 marginal increments over 
the standardized design region $\mathcal{{X}}=[0,1]^2$ is employed. 
The single point time plan is chosen as $t_1=\Delta_1=0.3$. The resulting optimal designs are derived in regards to the nominal values of parameters
are given in Table \ref{zuuzuzuz456546}, the normal use conditions $x_{u1}=-0.40$ and $x_{u2}=-0.60$, and the failure thresholds $z_{10}=2.56$ and $z_{20}=2.37$.


The D-optimal designs computed by the algorithm are the following four-point designs, which nearly coincide for the two copula,   
\begin{eqnarray*}
\mbox{Frank copula: }\   {\xi}_{_D}^* &=& \left(\begin{array}{ccccc}(0,0)&(0,1)&(1,0)&(1,1) \\ 0.24&0.24&0.26&0.26\end{array} \right);\\
\mbox{Gaussian copula: }\  {\xi}_{_D}^* &=& \left(\begin{array}{ccccc}(0,0)&(0,1)&(1,0)&(1,1) \\ 0.22&0.23&0.27&0.28\end{array} \right).
\end{eqnarray*}
The $c$-optimal designs from the algorithm are again four-point designs, which nearly coincide on the location of support points and the optimal weights of extremal points with some differences in the optimal weights of the two middle points,
\begin{eqnarray*}
\mbox{Frank copula: } \  {\xi}_{c}^* &=& \left(\begin{array}{ccccc}(0,0)&(0,1)&(0.5,1)&(1,1) \\ 0.09&0.18&0.46&0.27\end{array} \right);\\
\mbox{Gaussian copula: } \  {\xi}_{c}^* &=& \left(\begin{array}{ccccc}(0,0)&(0,1)&(0.5,1)&(1,1) \\ 0.11&0.22&0.39&0.28\end{array} \right).
\end{eqnarray*}
To evaluate the behaviour of the resulting optimal designs we consider the variations of the optimal weights when the underlying nominal values are misspecified. For brevity we consider the $c$-optimal design ${\xi}_{c}^*$ on the basis of the Gaussian copula function under deviations of the normal use condition $x_{u1}$, and the correlation parameter $\rho$. The four optimal weights $\omega_1^*$, $\omega_2^*$, $\omega_3^*$ and $\omega_4^*$ are plotted in Figure~\ref{weight-xu1-c-binary-gaussian} in dependence on $x_{u1}$ where all parameters are held fixed to their nominal values and in Figure~\ref{weight-rho-c-binary-gaussian} in dependence on $\rho$ where all parameters are held fixed to their nominal values. Figure~\ref{weight-xu1-c-binary-gaussian} shows that the optimal weights of the middle point, i.e. $w_2^*$ and $w_3^*$, considerably vary under changes of $x_{u1}$ where the optimal weights of the extermal point, i.e. $w_2^*$ and $w_3^*$, are nearly constant throughout. Figure~\ref{weight-rho-c-binary-gaussian} indicates that the resulting optimal design is more robust against misspecification of the correlation parameter $\rho.$
The nominal value for $x_{u1}$ and $\rho$ at $ {\xi}_{c}^*$ are indicated by vertical dotted lines in the corresponding figure.  
Define by
\[
	\mathrm{eff}(\xi) = \frac{\BS{c}^{T} \mathbf{M}(\xi_c^*,\BS{\beta})^{-1}\BS{c}}
{\BS{c}^{T} \mathbf{M}(\xi,\BS{\beta})^{-1}\BS{c}},
\]
the efficiency of of a design $\xi$ in terms of $\xi_c^*$ where $\BS{c}^{T} \mathbf{M}(\xi_c^*,\BS{\beta})^{-1}\BS{c}$ indicates the asymptotic variance for estimating $P_{11}$ under the optimal design $\xi_c^*$.
\begin{figure}
  \centering
  \begin{minipage}[b]{0.43\textwidth}
    \includegraphics[width=\textwidth]{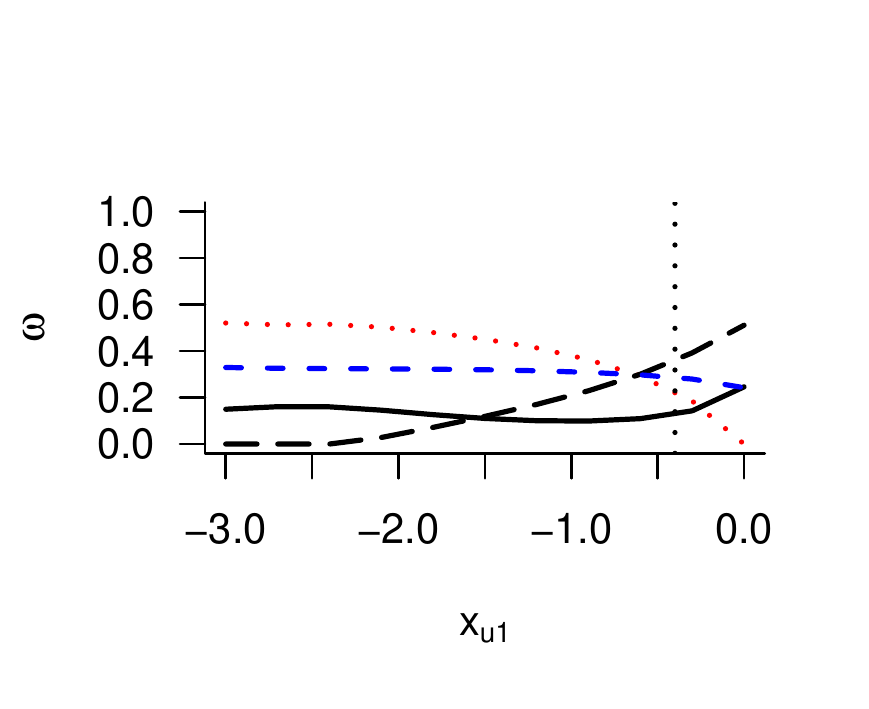}
    \caption{Dependence of the optimal weights on $x_{u1}$ for Example~\ref{sec:3.4656}, $\omega^*_1$: solid line, $\omega^*_2$: dotted line, $\omega^*_3$: long dashed line, $\omega^*_4$: dashed line }
\label{weight-xu1-c-binary-gaussian}
  \end{minipage}
  \hfill
  \begin{minipage}[b]{0.42\textwidth}
    \includegraphics[width=\textwidth]{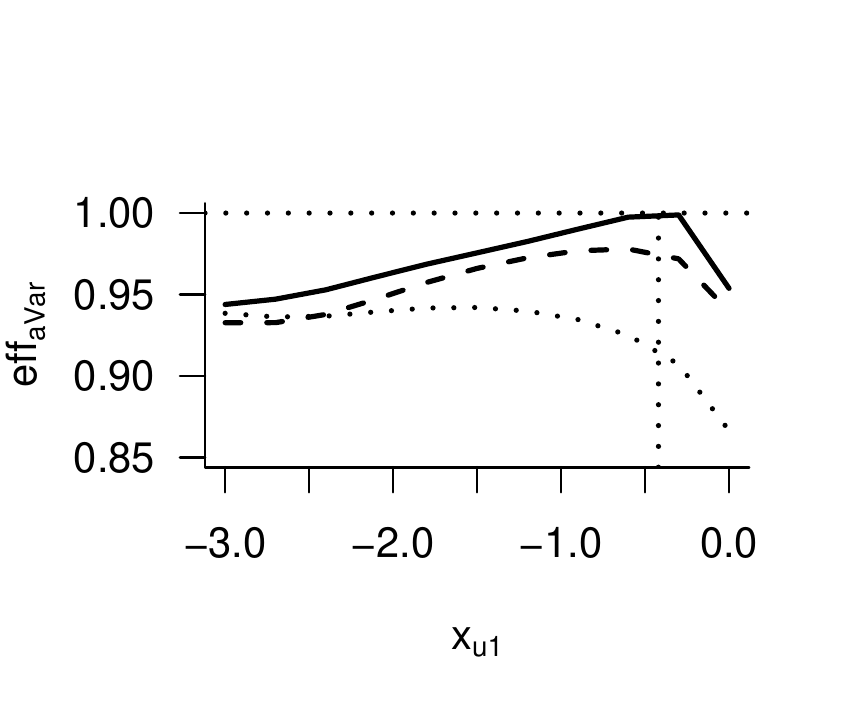}
    \caption{Efficiency of $\xi_{c}^*$ (solid line), $ \bar\xi_2$ (dashed line) and $ \bar\xi_3$ (dotted line) in dependence on $x_{u1}$ for Example~\ref{sec:3.4656}}
\label{eff-xu1-c-binary-gaussian}
  \end{minipage}
\end{figure}
\begin{figure}
  \centering
  \begin{minipage}[b]{0.43\textwidth}
    \includegraphics[width=\textwidth]{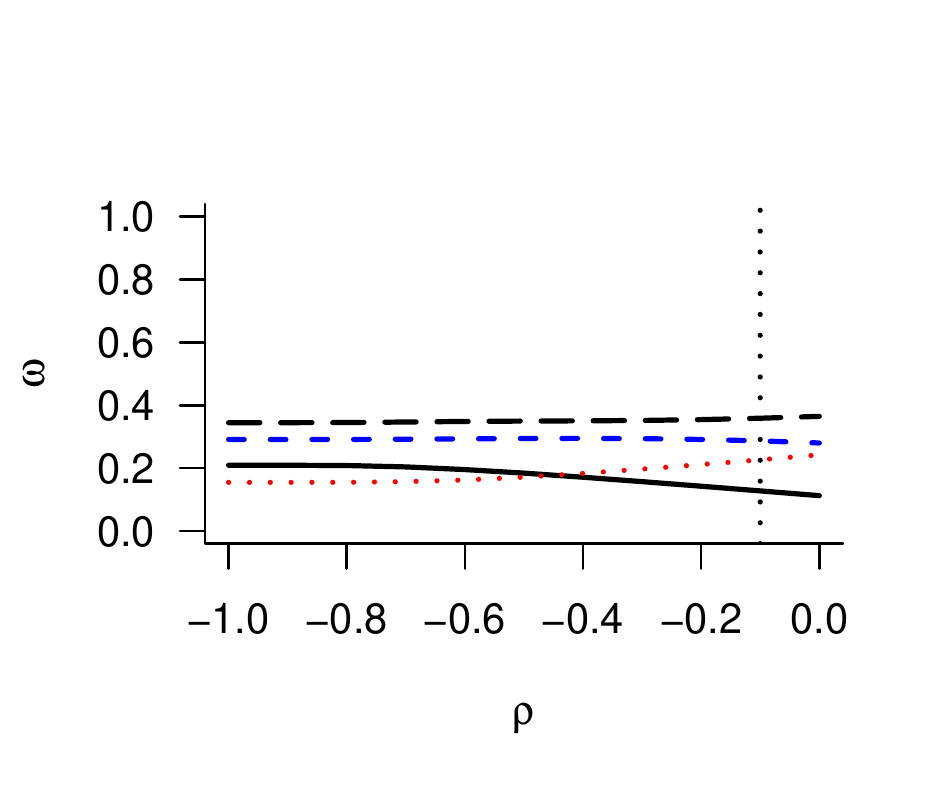}
    \caption{Dependence of the optimal weights on $\rho$ for Example~\ref{sec:3.4656}, $\omega^*_1$: solid line, $\omega^*_2$: dotted line, $\omega^*_3$: long dashed line, $\omega^*_4$: dashed line }
\label{weight-rho-c-binary-gaussian}
  \end{minipage}
  \hfill
  \begin{minipage}[b]{0.42\textwidth}
    \includegraphics[width=\textwidth]{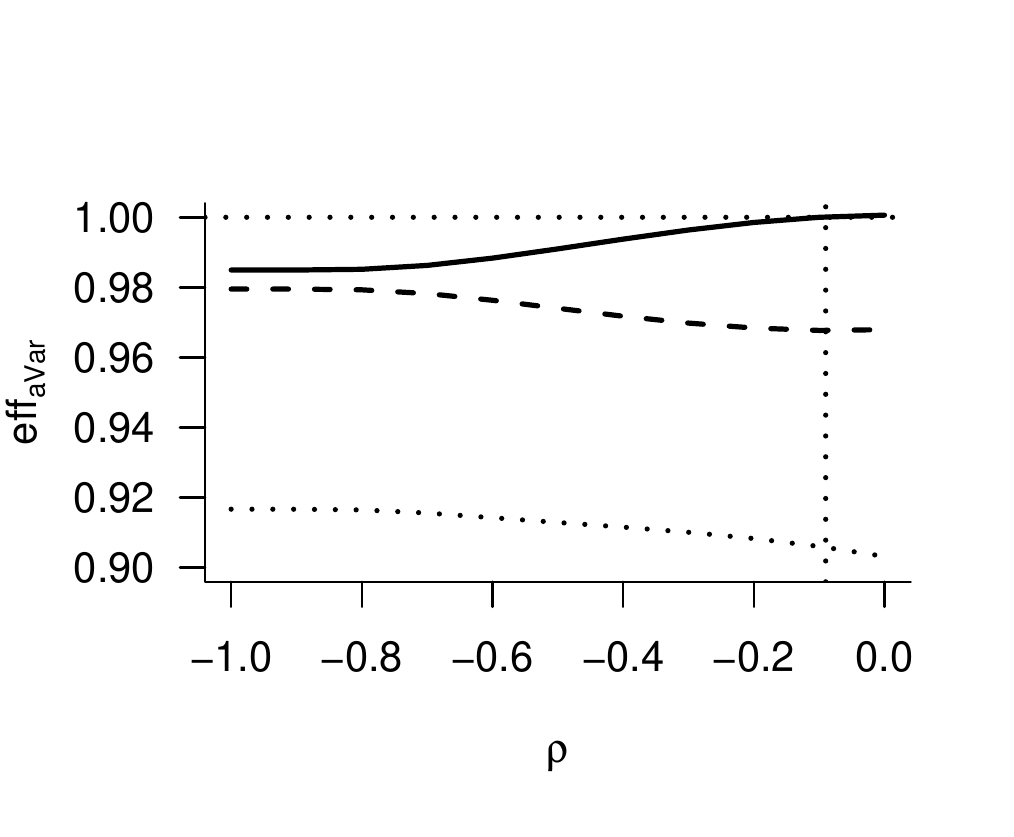}
    \caption{Efficiency of $\xi_{c}^*$ (solid line), $ \bar\xi_2$ (dashed line) and $ \bar\xi_3$ (dotted line) in dependence on $\rho$ for Example~\ref{sec:3.4656}}
\label{eff-rho-c-binary-gaussian}
  \end{minipage}
\end{figure}
Figure~\ref{eff-xu1-c-binary-gaussian} and Figure~\ref{eff-rho-c-binary-gaussian} show, respectively, the efficiencies in dependence on $x_{u1}$ and $\rho$  together with the efficiency of the $\bar\xi_2$ which assigns equal weights $1/4$ to the same support points of $ {\xi}_{c}^*$, and the design $\bar\xi_3$ which assigns equal weights $1/4$ to the vertices (0,0), (0,1), (1,0) and (1,1). 
Again, the nominal values for $x_{u1}$for $x_{u1}$ and $\rho$ at $ {\xi}_{c}^*$ are indicated by vertical dotted lines in the corresponding figure.  In total, Figure~\ref{eff-xu1-c-binary-gaussian} and Figure~\ref{eff-rho-c-binary-gaussian} indicate that the optimal design $\xi_{_c}^*$ performs quite well over the range of $x_{u1}$ and $\rho$ when compared to $\bar\xi_2$ and $\bar\xi_3$, which indicate that the optimal design is robust against changes of the normal use conditions as well as the nominal values. The existing results of the sensitivity analysis of ${\xi}_{c}^*$ on the basis of the Frank copula nearly coincides with the obtained results in regards to the Gaussian copula, and, hence, the latter results have been removed to avoid redundancy.
\end{example}

\section{Concluding remarks}
Reliability engineers are demanded to provide a sophisticated assessment of the reliability related properties during the design 
stage of highly reliable systems. 
Accelerated degradation testing (ADT) is a common approach to handle this issue. 
Accelerated degradation tests have the advantage to give an estimation of lifetime and reliability characteristics of the 
system under study in a relatively short testing time. 
In this work, we introduced optimal experimental designs for accelerated degradation tests with two response components 
and repeated measures with or without dependence between marginal components. The marginal degradation paths are expressed 
using Gamma process models. In the current models for ADT, we assume that stress remains constant 
within each unit during the whole test but may vary between units. Further, the same time plan for measurements is used 
for all units in the test.

In the case of independent components, it is desirable to estimate certain quantiles of the joint failure time distribution as a 
characteristic of the reliability of the product. Hence, the purpose of optimal experimental design is to find the best 
settings for the stress variable to obtain most accurate estimates of the quantiles.

On the other hand, the Frank copula as well as the Gaussian copula are separately adopted to represent the dependence 
relation in bivariate Gamma models when dependence is assumed between response components. 
The $D$-criterion is considered for locally optimal designs in both cases. The resulting optimal designs coincide in terms of 
the optimal support points but differ in their weights allocated to the points.

We developed further $D$- and $c$-optimal designs when the two Copula-based models are reduced to binary responses. 
A sensitivity analysis showed that the resulting locally optimal designs are quite efficient against deviations from the assumed nominal values.

Throughout, Gamma process models were considered as marginal degradation models. As a topic for future research, 
the results should be extended to other marginal failure models, e.\,g. 
Wiener process, inverse Gaussian process or non-linear mixed-effects degradation models.


\appendix 


 


\label{Appendix} 

\section{Derivation of the information matrix in Subsection \ref{information-copulaaaaa}}
\label{part444442343}
We derive formulas (\ref{elem-info-a}) by developing the double integral in (\ref{info-matrix-task}). Since the index $j$ will be fixed
in our derivations, we simply write $F_l,f_l,Y_l$ instead of  $F_{jl},f_{jl},Y_{jl}$, respectively, $l=1,2$. 
Recall the partitioning of $\BS{\beta}$ as
$\BS{\beta}=\bigl(\BS{\beta}_1^{T},\BS{\beta}_2^{T}\bigr)^{T}$, where $\BS{\beta}_l=\bigl(\beta_{1l},\beta_{2l}\bigr)^{T}$, $l=1,2$.
By $c_1(r,s)$, $c_2(r,s)$, $c_{11}(r,s)$, $c_{22}(r,s)$, and $c_{12}(r,s)$ we denote the partial derivatives of $c(r,s)$,
\begin{eqnarray*}
&&c_1(r,s)=\frac{\partial c(r,s)}{\partial r},\quad c_2(r,s)=\frac{\partial c(r,s)}{\partial s},\\
&&c_{11}(r,s)=\frac{\partial^2 c(r,s)}{\partial r^2},\quad c_{22}(r,s)=\frac{\partial^2 c(r,s)}{\partial s^2},\quad
c_{12}(r,s) = \frac{\partial^2 c(r,s)}{\partial r\,\partial s}.
\end{eqnarray*}
By straightforward calculation,
\[
\frac{\partial \ln c\bigl(F_1(Y_1),F_2(Y_2)\bigr)}{\partial\BS{\beta}_l} =
\frac{c_l\bigl(F(Y_1),F_2(Y_2)\bigr)}{c\bigl(F_1(Y_1),F_2(Y_2)\bigr)}\,\frac{\partial F_l(Y_l)}{\partial\BS{\beta}_l},\ \ l=1,2;
\]

\begin{eqnarray}
&&\frac{\partial^2\ln c\bigl(F_1(Y_1),F_2(Y_2)\bigr)}{\partial\BS{\beta}_l\BS{\beta}_l^{T}}= \label{eqA-0}\\
&&\phantom{xxxxxx}\left[\frac{c_{ll}\bigl(F_1(Y_1),F_2(Y_2)\bigr)}{c\bigl(F_1(Y_1),F_2(Y_2)\bigr)} - 
\frac{c_l^2\bigl(F_1(Y_1),F_2(Y_2)\bigr)}{c^2\bigl(F_1(Y_1),F_2(Y_2)\bigr)}\right]
\,\frac{\partial F_l(Y_l)}{\partial\BS{\beta}_l}\,
\Bigl(\frac{\partial F_l(Y_l)}{\partial\BS{\beta}_l}\Bigr)^{T} \nonumber\\ 
&&\phantom{xxxxxxxxxxxxxxxxxxxxxxx} +\,\frac{c_l\bigl(F_1(Y_1),F_2(Y_2)\bigr)}{c\bigl(F_1(Y_1),F_2(Y_2)\bigr)}\,
\frac{\partial^2F_l(Y_l)}{\partial\BS{\beta}_l\partial\BS{\beta}_l^{T}},\ \ \ \l=1,2;
\end{eqnarray}

\begin{eqnarray}
&&\frac{\partial^2\ln c\bigl(F_1(Y_1),F_2(Y_2)\bigr)}{\partial\BS{\beta}_1\partial\BS{\beta}_2^{T}} = 
\left[\frac{c_{12}\bigl(F_1(Y_1),F_2(Y_2)\bigr)}{c\bigl(F_1(Y_1),F_2(Y_2)\bigr)}\right. \nonumber\\ 
&&\phantom{xxxxxxxx} - \left. \frac{c_1\bigl(F_1(Y_1),F_2(Y_2)\bigr)\,c_2\bigl(F_1(Y_1),F_2(Y_2)\bigr)}
{c^2\bigl(F_1(Y_1),F_2(Y_2)\bigr)}\right]\,\frac{\partial F_1(Y_1)}{\partial\BS{\beta}_1}\,
\Bigl(\frac{\partial F_2(Y_2)}{\partial\BS{\beta}_2}\Bigr)^{T}.\label{eqA-0a}
\end{eqnarray}
We show that
\begin{equation}
{\rm E}\left(\frac{c_{ll}\bigl(F_1(Y_1),F_2(Y_2)\bigr)}{c\bigl(F_1(Y_1),F_2(Y_2)\bigr)}\,
\frac{\partial F_l(Y_l)}{\partial\BS{\beta}_l}\,\Bigl(\frac{\partial F_l(Y_l)}{\partial\BS{\beta}_l}\Bigr)^{T}\right)\,=\BS{0},
\ \ l=1,2.\label{eqA-1}
\end{equation}
Using the joint density of $\BS{Y}=(Y_1,Y_2)$ from (\ref{biv-df}), the expectation on the l.h.s. of (\ref{eqA-1}) rewrites, when $l=1$, as
\begin{eqnarray*}
&&\int_0^\infty\int_0^\infty c_{11}\bigl(F_1(y_1),F_2(y_2)\bigr)\,
\frac{\partial F_1(y_1)}{\partial\BS{\beta}_1}\,\Bigl(\frac{\partial F_1(y_1)}{\partial\BS{\beta}_1}\Bigr)^{T}
f_1(y_1)\,f_2(y_2)\,{\rm d}y_2{\rm d}y_2\\
&&= \int_0^\infty\frac{\partial F_1(y_1)}{\partial\BS{\beta}_1}\,\Bigl(\frac{\partial F_l(y_l)}{\partial\BS{\beta}_l}\Bigr)^{T}
f_1(y_1)\,\left\{\int_0^\infty c_{11}\bigl(F_1(y_1),F_2(y_2)\bigr)\,f_2(y_2)\,{\rm d}y_2\right\}\,{\rm d}y_1.
\end{eqnarray*}
For any fixed $y_1$, the inner integral becomes, by substituting $s=F_2(y_2)$ and interchanging integral and derivatives,  
\[
\int_0^1 c_{11}\bigl(F_1(y_1),s\bigr)\,{\rm d}s = \frac{\partial^2}{\partial r^2}\int_0^1 c(r,s)\,{\rm d}s\,\Big|_{r=F_1(y_1)}\,=0,
\]
where the last equation follows from $\int_0^1 c(r,s)\,{\rm d}s\,=1$ for all $0<r<1$. Hence (\ref{eqA-1}) follows for $l=1$,
and the case $l=2$ can be proved analoguously. 
Next we show that
\begin{equation}
{\rm E}\left(\frac{c_l\bigl(F_1(Y_1),F_2(Y_2)\bigr)}{c\bigl(F_1(Y_1),F_2(Y_2)\bigr)}\,
\frac{\partial^2F_l(Y_l)}{\partial\BS{\beta}_l\partial\BS{\beta}_l^{T}}\right)\,=\BS{0},\ \ \ l=1,2.\label{eqA-2}
\end{equation}
Again using the density from (\ref{biv-df}) and restricting to $l=1$ (the case $l=2$ is analogous), 
the expectation on the l.h.s. of (\ref{eqA-2}) rewrites as
\begin{eqnarray*}
&&\int_0^\infty\int_0^\infty c_1\bigl(F_1(Y_1),F_2(Y_2)\bigr)\,\frac{\partial^2 F_1(y_1)}{\partial\BS{\beta}_1\partial\BS{\beta}_1^{T}}\,
f_1(y_1)\,f_2(y_2)\,{\rm d}y_2{\rm d}y_1\\
&&=\int_0^\infty \frac{\partial^2 F_1(y_1)}{\partial\BS{\beta}_1\partial\BS{\beta}_1^{T}}\,f_1(y_1)
\left\{\int_0^\infty c_1\bigl(F_1(Y_1),F_2(Y_2)\bigr)\,f_2(y_2)\,{\rm d}y_2\right\}\,{\rm d}y_1,
\end{eqnarray*}
and for any fixed $y_1$ the inner integral is equal to
\[
\int_0^1c_1\bigl(F_1(y_1),s\bigr)\,{\rm d}s = \frac{\partial}{\partial r}\int_0^1 c(r,s)\,{\rm d}s\,\Big|_{r=F_1(y_1)}\,=0.
\]
From (\ref{eqA-1}), (\ref{eqA-2}), and (\ref{eqA-0}) it follows that 
\begin{eqnarray}
&&{\rm E}\left(\frac{\partial^2\ln c\bigl(F_1(Y_1),F_2(Y_2)\bigr)}{\partial\BS{\beta}_l\BS{\beta}_l^{T}}\right)=
{\rm E}\left(\frac{c_l^2\bigl(F_1(Y_1),F_2(Y_2)\bigr)}
{c^2\bigl(F_1(Y_1),F_2(Y_2)\bigr)}\,\frac{\partial F_l(Y_l)}{\partial\BS{\beta}_l}\,
\Bigl(\frac{\partial F_l(Y_l)}{\partial\BS{\beta}_l}\Bigr)^{T}\right)\nonumber\\
&&\phantom{x}-\int_0^\infty\int_0^\infty
\frac{c_l^2\bigl(F_1(Y_1),F_2(Y_2)\bigr)}
{c\bigl(F_1(Y_1),F_2(Y_2)\bigr)}\,\frac{\partial F_l(Y_l)}{\partial\BS{\beta}_l}\,
\Bigl(\frac{\partial F_l(Y_l)}{\partial\BS{\beta}_l}\Bigr)^{T} f_1(y_1)\,f_2(y_2)\,{\rm d}y_1{\rm d}y_2,\ \ l=1,2.
\label{eqA-3}
\end{eqnarray}
Next we show that
\begin{eqnarray}
&&{\rm E}\left(\frac{c_{12}\bigl(F_1(Y_1),F_2(Y_2)\bigr)}{c\bigl(F_1(Y_1),F_2(Y_2)\bigr)}\,\frac{\partial F_1(Y_1)}{\partial\BS{\beta}_1}\,
\Bigl(\frac{\partial F_2(Y_2)}{\partial\BS{\beta}_2}\Bigr)^{T}\right)\,= \label{eqA-4})\\
&&\phantom{xxxxxxxxx}\int_0^\infty\int_0^\infty c\bigl(F_1(y_1),F_2(y_2)\bigr)\,\frac{\partial f_1(y_1)}{\partial\BS{\beta}_1}\,
\Bigl(\frac{\partial f_2(y_2)}{\partial\BS{\beta}_2}\Bigr)^{T}{\rm d}y_1{\rm d}y_2.\nonumber
\end{eqnarray}
The expectaion on the l.h.s. of (\ref{eqA-4}) equals
\[
\int_0^\infty\int_0^\infty c_{12}\bigl(F_1(y_1),F_2(y_2)\bigr)\,
\frac{\partial F_1(y_1)}{\partial\BS{\beta}_1}\,
\Bigl(\frac{\partial F_2(y_2)}{\partial\BS{\beta}_2}\Bigr)^{T}\,f_1(y_1)\,f_2(y_2)\,{\rm d}y_1{\rm d}y_2.
\]
Writing
\[
c_{12}\bigl(F_1(y_1),F_2(y_2)\bigr)\,\frac{\partial F_1(y_1)}{\partial\BS{\beta}_1}\,f_1(y_1) = 
\frac{\partial}{\partial\BS{\beta}_1}\bigl[c_2\bigl(F_1(y_1),F_2(y_2)\bigr)\,f_1(y_1)\bigr]\,
-\,c_2\bigl(F_1(y_1),F_2(y_2)\bigr)\,\frac{\partial f_1(y_1)}{\partial\BS{\beta}_1},
\]
the last double integral rewrites as
\[
\int_0^\infty\left\{\int_0^\infty \Bigl[\frac{\partial}{\partial\BS{\beta}_1}\bigl[c_2\bigl(F_1(y_1),F_2(y_2)\bigr)\,f_1(y_1)\bigr]
- c_2\bigl(F_1(y_1),F_2(y_2)\bigr)\frac{\partial f_1(y_1)}{\partial\BS{\beta}_1}\Bigr]{\rm d}y_1\right\}
\Bigl(\frac{\partial F_2(y_2)}{\partial\BS{\beta}_2}\Bigr)^{T} f_2(y_2)\,{\rm d}y_2.
\]
Now for any fixed $y_2$,
\[
\int_0^\infty \frac{\partial}{\partial\BS{\beta}_1}\bigl[c_2\bigl(F_1(y_1),F_2(y_2)\bigr)\,f_1(y_1)\bigr]\,{\rm d}y_1 =
\frac{\partial}{\partial \BS{\beta}_1}\int_0^1c_2\bigl(r,F_2(y_2)\bigr)\,{\rm d}r\,=\BS{0},
\]
since the last integral does not depend on $\BS{\beta}_1$. We have obtained that the expectation  on the l.h.s. of (\ref{eqA-4})
is equal to
\begin{eqnarray*}
&&-\int_0^\infty\int_0^\infty c_2\bigl(F_1(y_1),F_2(y_2)\bigr)\,\frac{\partial f_1(y_1)}{\partial\BS{\beta}_1}\,
\Bigl(\frac{\partial F_2(y_2)}{\partial\BS{\beta}_2}\Bigr)^{T} f_2(y_2)\,{\rm d}y_1{\rm d}y_2\\
&&= - \int_0^\infty\frac{\partial f_1(y_1)}{\partial\BS{\beta}_1}\,\left\{\int_0^\infty
c_2\bigl(F_1(y_1),F_2(y_2)\bigr)\,\Bigl(\frac{\partial F_2(y_2)}{\partial\BS{\beta}_2}\Bigr)^{T} f_2(y_2)\,{\rm d}y_2\right\}
\,{\rm d}y_1.
\end{eqnarray*}
Writing
\begin{eqnarray*}
&&c_2\bigl(F_1(y_1),F_2(y_2)\bigr)\,\Bigl(\frac{\partial F_2(y_2)}{\partial\BS{\beta}_2}\Bigr)^{T} f_2(y_2) = \\
&&\phantom{xxxxxx}\Bigl(\frac{\partial}{\partial\BS{\beta}_2}\bigl[c\bigl(F_1(y_1),F_2(y_2)\bigr)\,f_2(y_2)\bigr]\Bigr)^{T}
-\,c\bigl(F_1(y_1),F_2(y_2)\bigr)\,\Bigl(\frac{\partial f_2(y_2)}{\partial\BS{\beta}_2}\Bigr)^{T},
\end{eqnarray*}
and observing that for any fixed $y_1$
\[
\int_0^\infty\frac{\partial}{\partial\BS{\beta}_2}\bigl[c\bigl(F_1(y_1),F_2(y_2)\bigr)\,f_2(y_2)\bigr]\,{\rm d}y_2 =
\frac{\partial}{\partial\BS{\beta}_2}\int_0^\infty c\bigl(F_1(y_1),s\bigr)\,{\rm d}s\,=\BS{0},
\]
we get (\ref{eqA-4}). From (\ref{eqA-0a}) and (\ref{eqA-4}) we get
\begin{eqnarray}
&&{\rm E}\left(\frac{\partial^2\ln c\bigl(F_1(Y_1),F_2(Y_2)\bigr)}{\partial\BS{\beta}_1\partial\BS{\beta}_2}\right) =
\label{eqA-5}\\
&&\phantom{x}\int_0^\infty\int_0^\infty
\Bigl[c\bigl(F_1(y_1),F_2(y_2)\bigr)\frac{\partial f_1(y_1)}{\partial\BS{\beta}_1}\,\frac{\partial f_2(y_2)}{\partial\BS{\beta}_2}
\nonumber\\
&&\phantom{xxxxxxx}
-\,\frac{c_1\bigl(F_1(y_1),F_2(y_2)\bigr)\,c_2\bigl(F_1(y_1),F_2(y_2)\bigr)}{c\bigl(F_1(y_1),F_2(y_2)\bigr)}\,
\frac{\partial F_1(y_1)}{\partial\BS{\beta}_1}\,\frac{\partial F_2(y_2)}{\partial\BS{\beta}_2}\,f_1(y_1)\,f_2(y_2)\Bigr]
\,{\rm d}y_1{\rm d}y_2.\nonumber
\end{eqnarray}
Observing that
\begin{eqnarray}
&&\frac{\partial F_l(y_l)}{\partial\BS{\beta}_l}=\frac{\partial F_l(y_l)}{\partial\gamma_l}\,
\frac{\partial\gamma_l}{\partial\BS{\beta}_l},\quad
\frac{\partial f_l(y_l)}{\partial\BS{\beta}_l}=\frac{\partial f_l(y_l)}{\partial\gamma_l}\,
\frac{\partial\gamma_l}{\partial\BS{\beta}_l},\label{eqA-6}\\
&&\mbox{ and }\ \ \frac{\partial\gamma_l}{\partial\BS{\beta}_l}=\gamma_l(x_l)\,\bigl(1,x_l\bigr)^{T},\ \ l=1,2,\label{eqA-7}
\end{eqnarray}
formulas (\ref{elem-info-a}) in Subsection \ref{information-copulaaaaa} follow from (\ref{eqA-3}) and (\ref{eqA-5}).\\

The derivatives $\frac{\partial f_l(y_l)}{\partial\gamma_l}$ and $\frac{\partial F_l(y_l)}{\partial\gamma_l}$ are given by 
\begin{eqnarray}
&&\frac{\partial}{\partial \gamma_l}f{_{l}}( {y}{_{l}})=\frac
{
\delta_{l}\big(\ln({y_{l}})-\ln(\nu_{_l})-\psi(\kappa_{l})\big)
}
{
{\Gamma(\kappa_{l})}\nu_{_l}^{\kappa_{l}}
},\\
\end{eqnarray}
where  $\kappa_{l}=\gamma_l\Delta$ and $\delta_{l}=
\exp\big({-{y}_{l}}/{\nu_{_l}}\big)    {y}_{{l}}^{\kappa_{l}-1}\Delta$, and

\begin{eqnarray}
\label{derivativeCDF22}
&&\frac{\partial }{\partial \gamma_l}F_{{ l}}(y_{{l}})=\frac{\partial }{\partial \gamma_l}\frac{\tilde\Gamma(\kappa_{l},y_{{l}}/\nu_l)}{\Gamma(\kappa_{l})}
=\Delta\bigg(-\Gamma(\kappa_{l}) ({y_{{l}}/\nu_l})^{\kappa_{l}}{}_2\tilde F_2(\kappa_{l},\kappa_{l};\kappa_{l}+1,\kappa_{l}+1;-{y_{l}/\nu_l})\\&&-\psi(\kappa_{l})\frac{\tilde\Gamma(\kappa_{l},{y_{{l}}/\nu_l})}{\Gamma(\kappa_{l})}-\exp({y_{{l}}/\nu_l})\frac{\Gamma(\kappa_{l},{y_{{l}}/\nu_l},0)}{\Gamma(\kappa_{l})}\bigg),\\
\end{eqnarray}

 such that $\psi(\kappa)=\frac{\partial}{\partial \kappa} \ln(\Gamma(\kappa))$ indicates the digamma function, $\Gamma(s,z,0)=\Gamma(s,z)-\Gamma(s)$, $\tilde\Gamma(\kappa_{l}, y_{{l}}/\nu_l)$ refers to the lower incomplete Gamma function, and ${}_{2}\tilde F_2$ denotes the regularized hypergeometric function which is extended from the generalized hypergeometric function ${}_2  F_2(\kappa,\kappa;\kappa+1,\kappa+1;-{y/\nu})$ and given by
\[
	{}_{2}\tilde F_2(\kappa, \kappa; \kappa + 1, \kappa + 1; - {y/\nu}) = \frac{
	1 + \sum_{k = 1}^\infty \left(\frac{\kappa}{\kappa + k}\right)^2 \frac{(- {y/\nu})^k}{k!}}{\Gamma (\kappa+1)^2}.
\]

\end{document}